\documentclass[aps,pra,twocolumn,showpacs,superscriptaddress,footinbib,10pt]{revtex4-1}
\usepackage{graphicx}
\usepackage{amsmath}

\newcommand{\be}{\begin{equation}}
\newcommand{\ee}{\end{equation}}
\newcommand{\BMPO}{Ba$_2$Mn(PO$_4$)$_2$}
\newcommand{\BNPO}{Ba$_2$Ni(PO$_4$)$_2$}

\usepackage{color}
\usepackage[colorlinks,breaklinks,bookmarks=true,citecolor=blue,linkcolor=red,urlcolor=blue]{hyperref}

\begin{document}

\title{Zigzag spin chains in the spin-5/2 antiferromagnet Ba$_2$Mn(PO$_4)_2$}

\author{Arvind Yogi}
\email{Present address: Center for Correlated Electron Systems, Institute for Basic Science (IBS), Seoul 08826, Korea; yogi.arvind2003@gmail.com}
\affiliation{Department of Condensed Matter Physics and Materials Science, Tata Institute of Fundamental Research, Homi Bhabha Road, Colaba, Mumbai 400 005, India}

\author{A. K. Bera}
\email{akbera@barc.gov.in}
\affiliation{Solid State Physics Division, Bhabha Atomic Research Centre, Mumbai 400085, India}

\author{Ashwin Mohan}
\affiliation{Institute of Chemical Technology, Nathalal Parekh Road, Matunga, Mumbai 400019, India}

\author{Ruta Kulkarni}
\affiliation{Department of Condensed Matter Physics and Materials Science, Tata Institute of Fundamental Research, Homi Bhabha Road, Colaba, Mumbai 400 005, India}

\author{S. M. Yusuf}
\affiliation{Solid State Physics Division, Bhabha Atomic Research Centre, Mumbai 400085, India}
\affiliation{Homi Bhabha National Institute, Anushaktinagar, Mumbai 400094, India}

\author{A. Hoser}
\affiliation{Helmholtz-Zentrum Berlin f\"{u}r Materialien und
Energie, 14109 Berlin, Germany}
%\email{hoser@helmholtz-berlin.de}

\author{A. A. Tsirlin}
\affiliation{Experimental Physics VI, Center for Electronic Correlations and Magnetism, Institute of Physics, University of Augsburg, 86135 Augsburg, Germany}

\author{A. Thamizhavel}
\email{thamiz@tifr.res.in}
\affiliation{Department of Condensed Matter Physics and Materials Science, Tata Institute of Fundamental Research, Homi Bhabha Road, Colaba, Mumbai 400 005, India}

\date{\today }

\begin{abstract}
Magnetic properties and magnetic structure of the Ba$_{2}$Mn(PO$_{4}$)$_{2}$ antiferromagnet featuring frustrated zigzag chains of $S=\frac{5}{2}$ Mn$^{2+}$ ions are reported based on neutron diffraction, density-functional band-structure calculations, as well as temperature- and field-dependent measurements of the magnetization and specific heat. A magnetic transition at $T_N\simeq 5$\,K marks the onset of the antiferromagnetic order with the propagation vector ${\mathbf k} = (\frac12\, 0\, \frac12)$ and ordered moment of $4.33\pm0.08~\mu_B$/Mn$^{2+}$ at 1.5\,K, pointing along the $c$ direction. Direction of the magnetic moment is chosen by the single-ion anisotropy, which is relatively weak compared to the isostructural Ni$^{2+}$ compound. Geometrical frustration has strong impact on thermodynamic properties of Ba$_2$Mn(PO$_4)_2$, but manifestations of the frustration are different from those in Ba$_2$Ni(PO$_4)_2$, where frustration by isotropic exchange couplings is minor, yet strong and competing single-ion anisotropies are present. A spin-flop transition is observed around 2.5\,T.  The evaluation of the magnetic structure from the ground state via the spin-flop state to the field-polarized ferromagnetic state has been revealed by a comprehensive neutron diffraction study as a function of magnetic field below $T_N$. Finally, a magnetic phase diagram in the $H-T$ plane is obtained.
\end{abstract}

%\keywords{Honeycomb lattice, Transition metal oxides, Neutron diffraction}

\maketitle
\section{Introduction}
\label{intro}
Different transition-metal ions embedded in the same type of crystal structure can show drastically different magnetic behavior. Even if structural arrangement does not change significantly, magnetic interactions can be altered because different magnetic orbitals become active~\cite{Kanamori1959,Anderson1959,Goodenough}. Additionally, the type of local (single-ion) anisotropy changes from one transition-metal ion to another. For example, Li$_2$CuW$_2$O$_8$ is collinear antiferromagnet with a single magnetic transition~\cite{Ranjith2015}, whereas Li$_2$NiW$_2$O$_8$ shows two ordered states in zero field, one commensurate and one incommensurate~\cite{Ranjith2016}. The consecutive formation of two ordered states may be driven by the single-ion anisotropy of spin-1 Ni$^{2+}$, which is absent in the case of spin-$\frac12$ Cu$^{2+}$. Despite the absence of anisotropy, Cu$^{2+}$ is a complex magnetic ion on its own with a strong proclivity to long-range superexchange couplings. Whereas BiMn$_2$PO$_6$ develops conventional long-range magnetic order~\cite{Nath2014}, BiCu$_2$PO$_6$ lacks magnetic order and exhibits gapped singlet ground state~\cite{Koteswararao2007,Mentre2009}. This is due to the second-neighbor couplings that introduce strong frustration~\cite{Mentre2009,Tsirlin2010} reflected by the non-trivial dispersion of triplet excitations~\cite{Plumb2013,Plumb2016} and a plethora of field-induced states~\cite{Kohama2012,Kohama2014}.

This variability and diversity can also be utilized to distinguish between local and cooperative effects, thus elucidating the microscopic nature of the system. Recently, we studied Ba$_2$Ni(PO$_4)_2$~\cite{Yogi2017}, where two consecutive magnetic transitions take place in zero field, a short-range order is observed above the magnetic transitions, and several field-induced phases are formed. Remarkably, this complex behavior occurs in the presence of mostly non-frustrated exchange couplings. However, single-ion anisotropy may be quite strong and leads to different easy axes for different Ni$^{2+}$ ions, thus frustrating collinear order imposed by the exchange.

Here, we seek to verify this scenario by studying the reference Mn$^{2+}$ compound, Ba$_2$Mn(PO$_4)_2$, with the same crystal structure but different magnetic ion. Whereas Ni$^{2+}$ ($d^8$) can support a sizable single-ion anisotropy, Mn$^{2+}$ should be nearly isotropic. By comparing the low-temperature magnetic behavior of the Mn and Ni compounds, we are able to distinguish between effects driven by the exchange and single-ion anisotropy. Indeed, we find that Ba$_2$Mn(PO$_4)_2$ shows neither multiple magnetic transitions nor short-range order, and its field-induced state is merely a spin-flop phase. This happens despite the stronger frustration of isotropic exchange couplings in Ba$_2$Mn(PO$_4)_2$ compared to its Ni-based sibling.

\section{Methods}
\label{sec:methods}
Polycrystalline samples of Ba$_{2}$Mn(PO$_{4}$)$_{2}$ were prepared by the solid-state reaction technique using BaCO$_{3}$ ($99.999$\%, Alfa Aesar), MnO ($99.999$\%, Aldich), and NH$_4$H$_2$PO$_4$ ($99.999$\%, Alfa Aesar) as starting materials. The stoichiometric mixtures were heated at $1000$\,$^{\circ}$C for 72 h with several intermediate grindings and pelletizations. The resulting sample had light pink color. Its phase purity was confirmed by x-ray diffraction (XRD, PANalytical powder diffractometer and CuK$_{\alpha}$ radiation) at room temperature. However, magnetization measurements revealed a minor amount of the ferromagnetic impurity Mn$_3$O$_4$ that could not be detected by XRD and neutron diffraction. The crystal structure was refined by the Rietveld method using the powder XRD data.

Magnetic susceptibility $\chi(T)$ measurements were performed using the SQUID magnetometer (Quantum Design). Heat capacity $C_p(T)$ measurements were performed using the commercial Physical Property Measurement System (PPMS, Quantum Design).

Neutron diffraction patterns were recorded using the cold neutron focusing diffractometer E6 ($\lambda$ = 2.45 {\AA}) at HZB, Germany. The powder sample was packed in a cylindrical vanadium container. Zero-field low-temperature measurements were performed in a standard orange He cryostat. The neutron diffraction patterns under magnetic fields were recorded by using a 14~T vertical cryomagnet from Oxford instruments. The diffraction data were analyzed by the Rietveld method using the \texttt{Fullprof}~\cite{fullprof}.

Isotropic exchange couplings were parametrized by the spin Hamiltonian
\begin{equation}
 H=\sum_{\langle ij\rangle}J_{ij}\,\mathbf S_i\mathbf S_j-\sum_iD\,(S_i^z)^2+\sum_iE[(S_i^x)^2-(S_i^y)^2],
\label{eq:hamiltonian}\end{equation}
where the summation is over bonds $\langle ij\rangle$, $J_{ij}$ are isotropic exchange couplings, $D$ and $E$ stand, respectively, for the out-of-plane and in-plane single-ion anisotropies, and $S=\frac52$ for Mn$^{2+}$. Microscopic parameters entering Eq.~\eqref{eq:hamiltonian} were computed by density-functional (DFT) band-structure calculations performed in the \texttt{FPLO}~\cite{fplo} and \texttt{VASP}~\cite{vasp1,*vasp2} codes using the Perdew-Burke-Ernzerhof flavor of the exchange-correlation potential~\cite{pbe96}. Correlation effects in the Mn $3d$ shell were taken into account on the mean-field DFT+$U$ level with the on-site Coulomb repulsion parameter $U_d=5.5$\,eV and Hund's exchange $J_d=1$\,eV~\cite{Nath2014}. Magnetic parameters were obtained by a mapping procedure, as described in Ref.~\onlinecite{Xiang2013}. The local moment of 4.65\,$\mu_B$ on Mn was obtained in \texttt{VASP} calculations.

Thermodynamic properties for the isotropic part of the spin Hamiltonian, Eq.~\eqref{eq:hamiltonian}, were obtained by quantum Monte-Carlo simulations using the \texttt{loop} algorithm~\cite{loop} implemented in the \texttt{ALPS} package~\cite{alps}. $L\times L\times L$ finite lattices ($L\leq 10$) with periodic boundary conditions were employed, excluding any appreciable finite-size effects within the temperature range of our interest. N\'eel temperatures were obtained as the crossing point of Binder cumulants for the staggered magnetization calculated with different $L$.

\section{Results}
\label{sec:results}
\subsection{{X-ray diffraction and crystal structure}
\label{sec:x-ray}}
\begin{figure}
\includegraphics[width=1.00\linewidth]{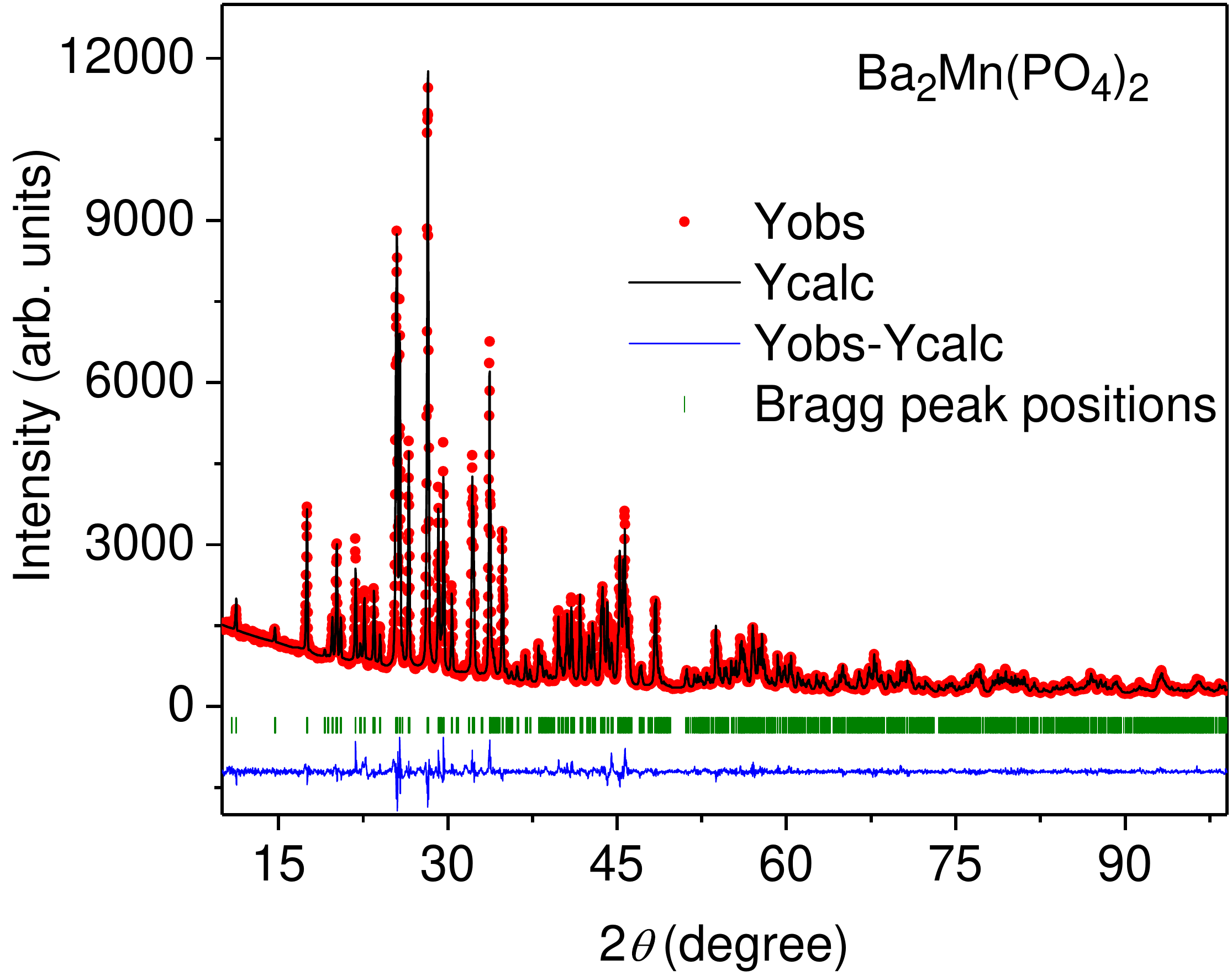}
                %\linewidth or \textwidth
 \caption{\label{XRD} (Color online) Rietveld refined x-ray powder diffraction pattern for Ba$_{2}$Mn(PO$_{4}$)$_{2}$ at room temperature. The observed and calculated patterns are shown by filled circles and solid black line, respectively. The difference between observed and calculated patterns is shown by the thin line at the bottom. The vertical bars are the allowed Bragg peak positions.}
\end{figure}

Rietveld refinement of the room-temperature XRD pattern (Fig.~\ref{XRD}) confirms that Ba$_2$Mn(PO$_4)_2$ crystallizes in the monoclinic space group $P2_1/n$ (space group No. 14). Refined lattice parameters (Table~\ref{tab:latticeparameters}) are in good agreement with the values reported by Faza \emph{et al}.~\cite{Faza2001}. There is a single crystallographic site for magnetic Mn, however, two sites for each of Ba and P ions. Oxygen ions are located at eight crystallographic sites. All atoms occupy at the 4\emph{e} crystallographic site and all the atomic sites are found to be fully occupied.

\begin{figure*}
\includegraphics[width=0.88\linewidth]{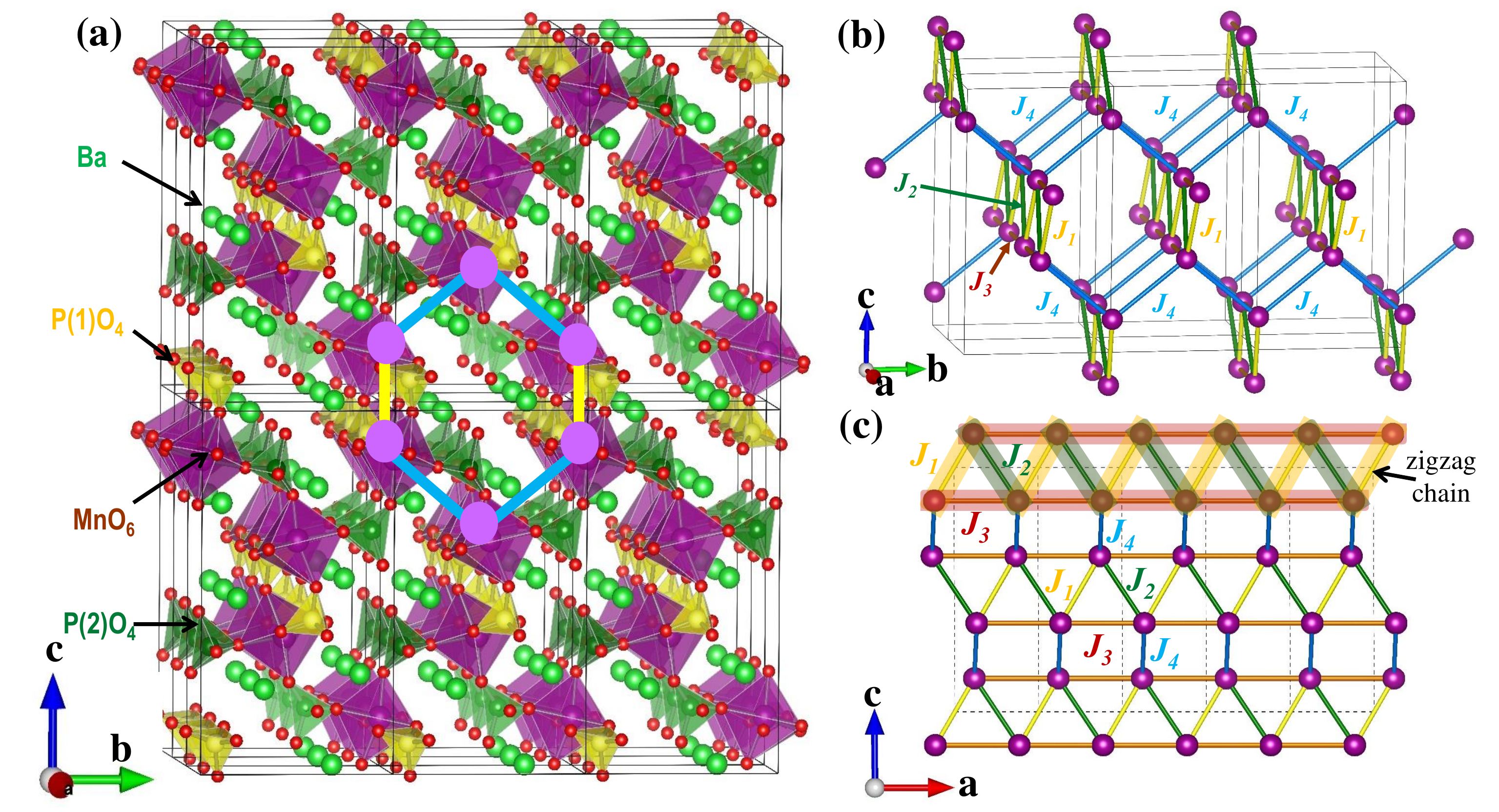}
  %\linewidth or \textwidth
\caption{\label{fig:structure}
(Color online) (a) Projection of the crystal structure of Ba$_{2}$Mn(PO$_{4}$)$_{2}$ along the $a$ axis, with the honeycomb arrangement of the Mn$^{2+}$ ions highlighted by the honeycomb-unit. (b) Schematic representation of the exchange interactions $J_1$, $J_2$, $J_3$ and $J_4$ in Ba$_{2}$Mn(PO$_{4}$)$_{2}$. The honeycomb units are formed by two $J_1$ and four $J_4$ exchange interactions, whereas $J_2$ and $J_3$ connect the honeycomb planes.(c) An alternative view of the spin lattice in terms of zigzag spin chains along the crystallographic $a$-axis. The chains are built by nearest-neighbor couplings $J_1$, $J_2$ and the second-neighbor coupling $J_3$. The coupling $J_4$ connects the zigzag chains.}
\end{figure*}

\begin{table}[ptb]
\caption{\label{tab:latticeparameters}
Refined fractional atomic coordinates and isotropic thermal parameters $B_{iso}$ of \BMPO\ at room temperature. Same value for $B_{\rm iso}$ parameter was used for atoms of each type. The numbers in the parentheses are error bars obtained from the Rietveld refinement. Lattice parameters are $a = 5.3160(9)$\,\r A, $b = 8.9750(15)$\,\r A, $c = 16.2750(3)$\,\r A, and $\beta$ = 90.241(9)$^{\circ}$. Space group: $P2_1/n$.}
\begin{ruledtabular}
\begin{tabular}{cccccccc}
Atom &$x/a$ & $y/b$ & $z/c$ & $B_{iso}$ (\AA$^2$)  \\\hline
Ba1 & $0.2897(4)$ & $0.2973(17)$ & $0.9800(11)$ & $0.69(04)$  \\
Ba2 &  $0.7373(4)$ & $0.8437(17)$ & $0.2705(11)$ & $0.69(04)$  \\
Mn & $0.2651(14)$ & $0.4845(5)$ & $0.3626(3)$  & $0.44(08)$  \\
P1 & $0.7480(2)$ & $0.5926(8)$ & $0.4194(5)$  & $0.89(11)$  \\
P2 & $0.2410(2)$ & $0.5972(8)$ & $0.1650(5)$  & $0.89(11)$ \\
O11 & $0.7820(4)$ & $0.6089(14)$ & $0.5109(10)$ & $0.99(19)$ \\
O12 & $0.5150(3)$ & $0.6880(2)$ & $0.4063(11)$ & $0.99(19)$  \\
O13 & $0.6940(4)$ & $ 0.4370(17)$ & $0.3851(9)$ & $0.99(19)$  \\
O14 & $0.9740(3)$ & $0.6570(2)$ & $0.3802(11)$ & $0.99(19)$  \\
O21 & $0.3130(3)$ & $0.5436(16)$ & $0.0716(10)$ & $0.99(19)$ \\
O22 & $0.3270(3)$ & $0.7537(18)$ & $0.1630(9)$ & $0.99(19)$ \\
O23 & $-0.0660(3)$ & $0.6111(20)$ & $0.1730(10)$ & $0.99(19)$ \\
O24 & $0.3080(4)$ & $0.5094(16)$ & $0.2392(10)$ & $0.99(19)$ \\
\end{tabular}
\end{ruledtabular}
\end{table}

The crystal structure of Ba$_{2}$Mn(PO$_{4}$)$_{2}$ consists of layers of MnO$_6$ octahedra and (PO$_4$)$^{-3}$ groups. All the magnetic Mn$^{2+}$ ions are equivalent by symmetry and situated at the center of a nearly regular MnO$_6$ octahedron. On the other hand, phosphorous atoms have two nonequivalent crystallographic sites, P(1) and P(2), that form nearly regular P(1)O$_{4}$ and P(2)O$_{4}$ tetrahedra, respectively. Within a given layer, the MnO$_6$ octahedra are connected by corner-sharing PO$_4$ tetrahedra, resulting in a honeycomb arrangement of the magnetic Mn$^{2+}$ ions [Figs.~\ref{fig:structure}(a) and \ref{fig:structure}(b)]. One of such honeycomb units of the Mn$^{2+}$ ions is shown in [Fig. \ref{fig:structure} (a)]. The honeycombs are distorted, with two shorter ($J_1$; 5.089 \r A) and four longer ($J_4$; 5.807 \r A) Mn--Mn contacts via the P(1)O$_4$ and P(2)O$_4$ tetrahedra, respectively [Fig. \ref{fig:structure} (b)]. Two other possible interactions ($J_2$ and $J_3$ between the Mn ions having distances 5.286 and  5.311 \r A, respectively, via the P(1)O$_4$ and P(2)O$_4$ tetrahedra) connect the adjacent honeycomb planes. The crystal structure of this compound can alternatively be viewed as a zigzag spin chain along the crystallographic $a$-axis [Fig. \ref{fig:structure} (c)]. The zigzag spin chains of Mn$^{2+}$ are formed by nearest-neighbor intrachain interactions $J_1$ and $J_2$. The finite second nearest-neighbor exchange interaction $J_3$ also possible due to the similar superexchnage interaction pathways. Such zigzag chains are again coupled by $J_4$ along the $c$-axis. Within the zigzag chains, the triangular arrangement of the three possible intrachain interactions ($J_1$-$J_3$) may lead to a geometrical frustration. We will show below that such a description is indeed more reasonable, and will also assess the effect of frustration on thermodynamic properties of Ba$_{2}$Mn(PO$_{4}$)$_{2}$.

%The crystal structure of this compound can alternatively be viewed as a zigzag spin chain along the crystallographic $a$-axis [Fig. \ref{fig:structure} (c)]. The linear chains of Mn ions with intra-chain interaction $J_3$ are coupled by nearest-neighbor couplings $J_1$ and $J_2$ along the $c$-axis to form the zigzag spin chain structure. Such nearest-neighbor couplings $J_1$, $J_2$ and the second-neighbor coupling $J_3$ are competing in nature due to the triangular geometry. This leads to the possibility of inter-chain coupling introducing frustration into the system. When viewed along the chain direction, the zigzag chains form a stacked-honeycomb lattice through a very week couplings $J_4$ as shown in Fig. \ref{fig:structure} (b).

\subsection{Magnetic Susceptibility}
\label{sec:magnetization}

\begin{figure}
\includegraphics[width=0.90\linewidth]{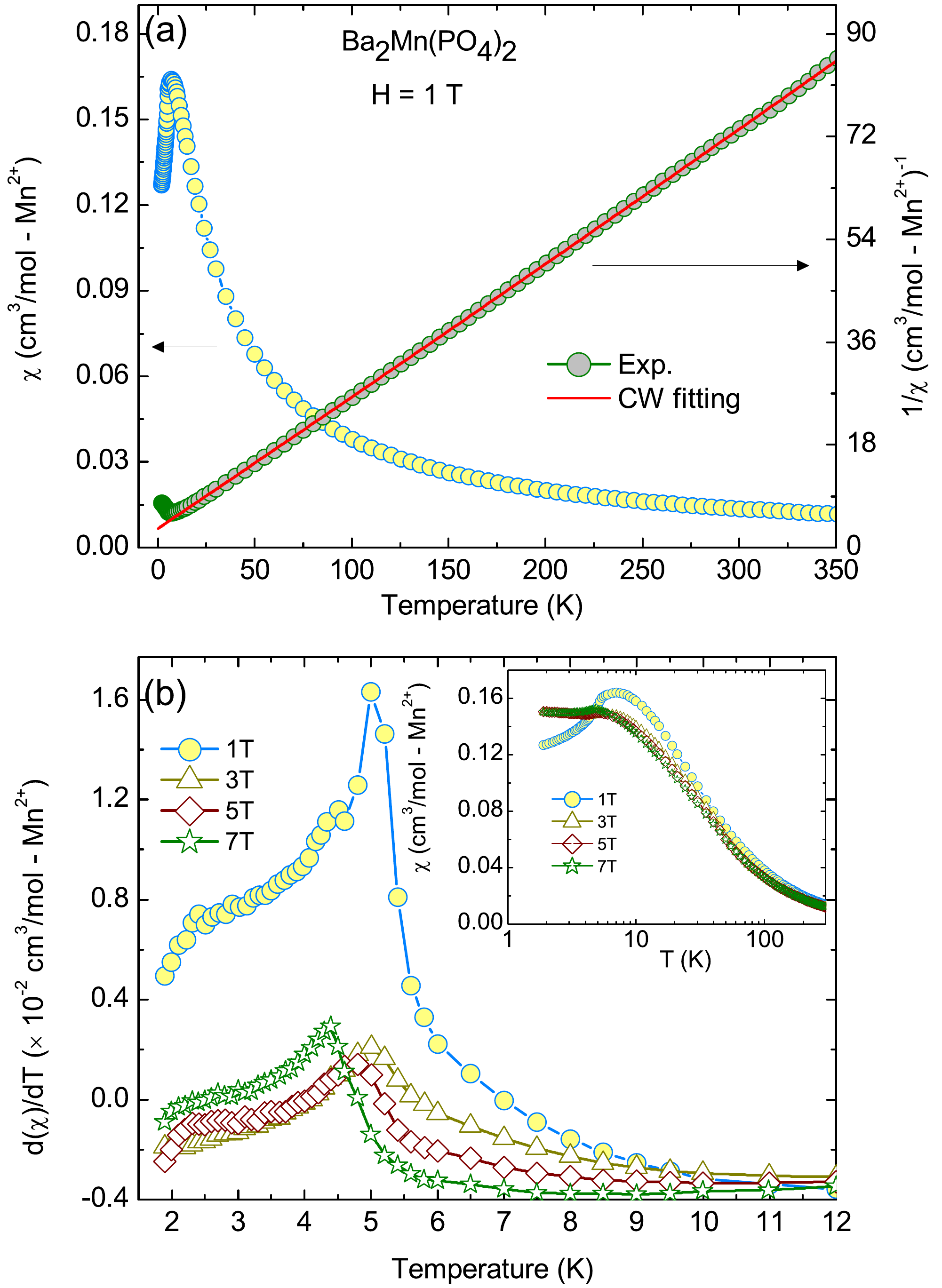}
                %\linewidth or \textwidth
\caption{\label{fig:CW}
(Color online) (a) Temperature-dependent dc-magnetic susceptibility ($\chi=M/H$) and the inverse susceptibility $\chi^{-1}$ of Ba$_{2}$Mn(PO$_{4}$)$_{2}$ measured under an applied magnetic field of $H=1$~T. The straight solid red line is the Curie-Weiss fit to the $\chi^{-1}(T)$ data above 15~K. (b) Fisher's heat capacity $\partial(\chi)/\partial T$ at different applied fields over the low-$T$ region. The inset shows $\chi(T)$ measured under different applied fields.}
\end{figure}

Temperature-dependent dc-magnetic susceptibility $\chi=M/H$ (Fig.~\ref{fig:CW}) reveals a kink at $T_N\simeq 5$\,K, which is especially pronounced in the Fisher's heat capacity $\partial(\chi T)/\partial T$ [Fig. \ref{fig:CW} (b)], suggesting the onset of long-range magnetic ordering, in agreement with heat capacity data discussed below. Additionally, weak ferromagnetic signal was observed in the field of 0.1\,T below 42\,K. We ascribe this ferromagnetic signal to a minor impurity of Mn$_3$O$_4$~\cite{Dwight1960}. Its contribution is fully suppressed already at 1\,T, no anomaly at 42\,K is seen.

At high temperatures, the susceptibility follows the Curie-Weiss behavior [Fig. \ref{fig:CW} (a)],
\begin{equation}
\chi=\frac{C}{T-\theta_{\rm CW}},
\label{Curie}
\end{equation}
where $C=N_A\mu_{\rm eff}^{2}/3k_B$ is the Curie constant, and $\theta_{\rm CW}$ is the Curie-Weiss temperature, $N_{A}$ stands for Avogadro's number, $\mu_{\rm eff}$ is the effective magnetic moment, and $k_{B}$ is Boltzmann's constant. The fit of Eq.~\eqref{Curie} to the experimental $\chi^{-1}(T)$ data above 80\,K yields $C = 4.26(1)$~cm$^{3}$~K/mol~Mn, and $\theta_{\rm CW} = -12.0(1)$\,K. The effective magnetic moment ($\mu_{\rm eff}$) is calculated to be 5.84(1)\,$\mu_{B}$/Mn$^{2+}$ in good agreement with the earlier report~\cite{Faza2001} and with the expected spin-only value of 5.92\,$\mu_{B}$ for $S = \frac{5}{2}$ and $g=2$. The negative Weiss temperature suggests that the dominating exchange interactions are antiferromagnetic (AFM) in nature. The ratio $|\theta_{\rm CW}$/$T_N|$~=~2.4 suggests the possibility of frustrations in this system. The $\chi(T)$ curves, measured under different applied magnetic fields, are shown in the inset of Fig. \ref{fig:CW}(b). With increasing the field, $T_N$ shifts toward lower temperatures.

\subsection{Heat Capacity}
\label{sec:heat-capacity}

\begin{figure}
\includegraphics[width=1\linewidth]{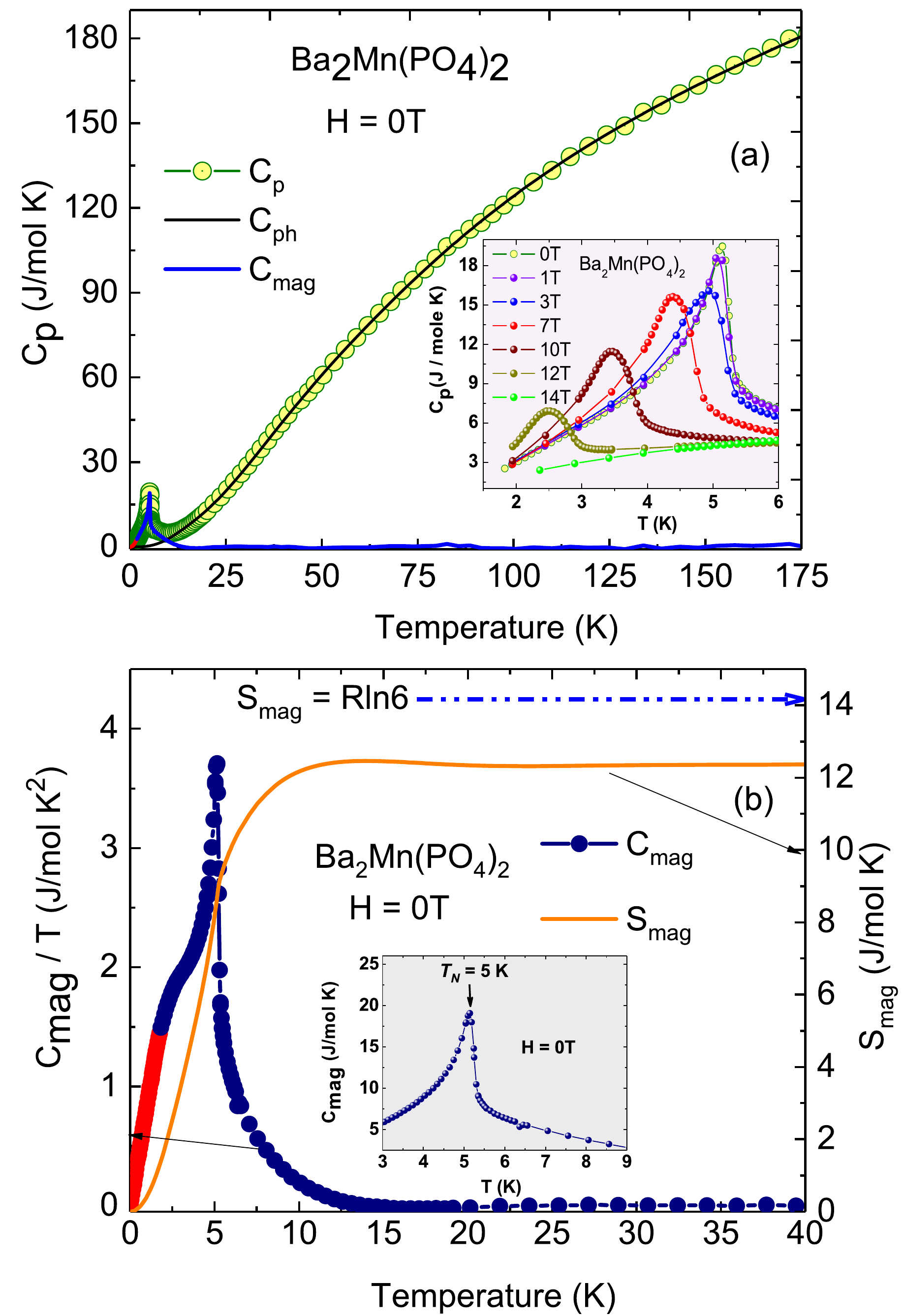}
\caption{\label{cp-cph}
(Color online) (a) Temperature dependence of the specific heat $C_{p}$ of Ba$_{2}$Mn(PO$_{4}$)$_{2}$ measured at zero field. Open circles are the raw data, the solid black line shows the phonon contribution $C_{\rm ph}$, according to the fit with Eq.~\eqref{DebyeE}, and the solid blue line indicates the magnetic contribution ($C_{\rm mag}$). The red solid line is extrapolated $C_{\rm mag}$ down to 0\,K. (b) The $C_{\rm mag}/T$ as a function of temperature over the low-temperature range ($1.8-40$\,K). The derived magnetic entropy ($S_{\rm mag}$) is shown by the solid orange line. The inset shows the $C_{\rm mag}$ curve zoomed over 3-9\,K.}
\end{figure}

Specific heat of Ba$_{2}$Mn(PO$_{4}$)$_{2}$ (Fig.~\ref{cp-cph}) reveals a $\lambda$-like anomaly at $T_N\simeq 5$\,K in zero field, confirming the onset of long-range magnetic ordering. A single magnetic phase transition is evident for the studied Mn compound in contrast to its Ni counterpart having two successive magnetic phase transition at 5 and 4.6 K. We estimated magnetic heat capacity, $C_{\rm mag}(T)$, by subtracting the lattice contribution from the measured total heat capacity data. The lattice contribution is estimated by fitting the high-temperature data above 20\,K  by a model with one Debye-type and three Einstein-type heat capacity terms,
\begin{equation}
C_p(T) = f_{\rm D}\,C_{\rm Deb} (\Theta_{\rm D},T) + \sum_i {g_i\,C_{{\rm Ein},i}(\Theta_{{\rm E},i},T)}.
\label{DebyeE}
\end{equation}

The detailed methodology of the fitting is explained in Ref.~\onlinecite{Yogi2017,Kittel,Koteswararao2014,Yogi2015}. The best fit of the data ($20-200$\,K) was obtained for $\Theta_{D}=85.53(8)$\,K, $\Theta_{E1}=166.21(4)$\,K, $\Theta_{E2}=399.36(4)$\,K, and $\Theta_{E3}=1074.03(5)$\,K. The fitted curve is shown by the solid line in Fig.~\ref{cp-cph}(a).
The fits of the experimental $C_{p}$ data by different combinations of the Debye and Einstein terms over various temperature ranges confirm that below 13\,K $C_{\rm mag}$ is the dominant contribution independent of fitting details.

This magnetic contribution is shown in Fig.~\ref{cp-cph}(b). The red line in Fig.~\ref{cp-cph}(a) and solid circles in Fig.~\ref{cp-cph}(b) below 1.8 K are data points extrapolated to 0\,K by assuming a simplified magnon dispersion of a 3D antiferromagnet ($C_{\rm mag}$~$\propto$~$T^{3}$), to estimate magnetic entropy ($S_{\rm mag}$) accurately.  The associated magnetic entropy is calculated as

\begin{equation}
S_{\rm mag}(T)=\int_0^T C_{\rm mag}(T')/T' dT'.
\label{EqS}
\end{equation}

The estimated $S_{\rm mag}$ as function of temperature is shown in Fig.~\ref{cp-cph}(b) by the solid orange line. The total amount of magnetic entropy change is $S_{\rm mag}$ = 12.36(2)~J/(mol K) at 40\,K, in good agreement (~83~$\%$) with the expected value of $R\ln(2S+1)= R\ln(6) = 14.89$~J/mol\,K, where $R$ is the molar gas constant.

Two features of the $C_{\rm mag}/T$ curve are worth mentioning. First, the broad bend seen below $T_N$ is typical of the specific heat of a spin-$\frac52$ magnetic ion and originates from temperature-dependent populations of the Zeeman levels split by the local exchange field~\cite{Johnston2011,Nath2014}. Second, no broad peak/hump is seen above $T_N$ [inset of Fig.~\ref{cp-cph}(b)], thus indicating the absence of short-range order and distinguishing Ba$_2$Mn(PO$_4)_2$ from its Ni sibling that shows a broad peak/hump in $C_{\rm mag}/T$ around 7\,K~\cite{Yogi2017}, i.e., 2.5\,K above the $T_N$~\cite{note3}.

The $C_p(T)$ curves measured under different applied magnetic fields are shown in the inset of Fig.~\ref{cp-cph}(a). With increasing the field, the $\lambda$-like anomaly broadens and shifts towards lower temperatures. Eventually, at 14\,T no transition is observed indicating that $T_N$ is suppressed below 1.8\,K or even vanishes.

\subsection{Magnetization}

\begin{figure}
\includegraphics[width=1.0\linewidth]{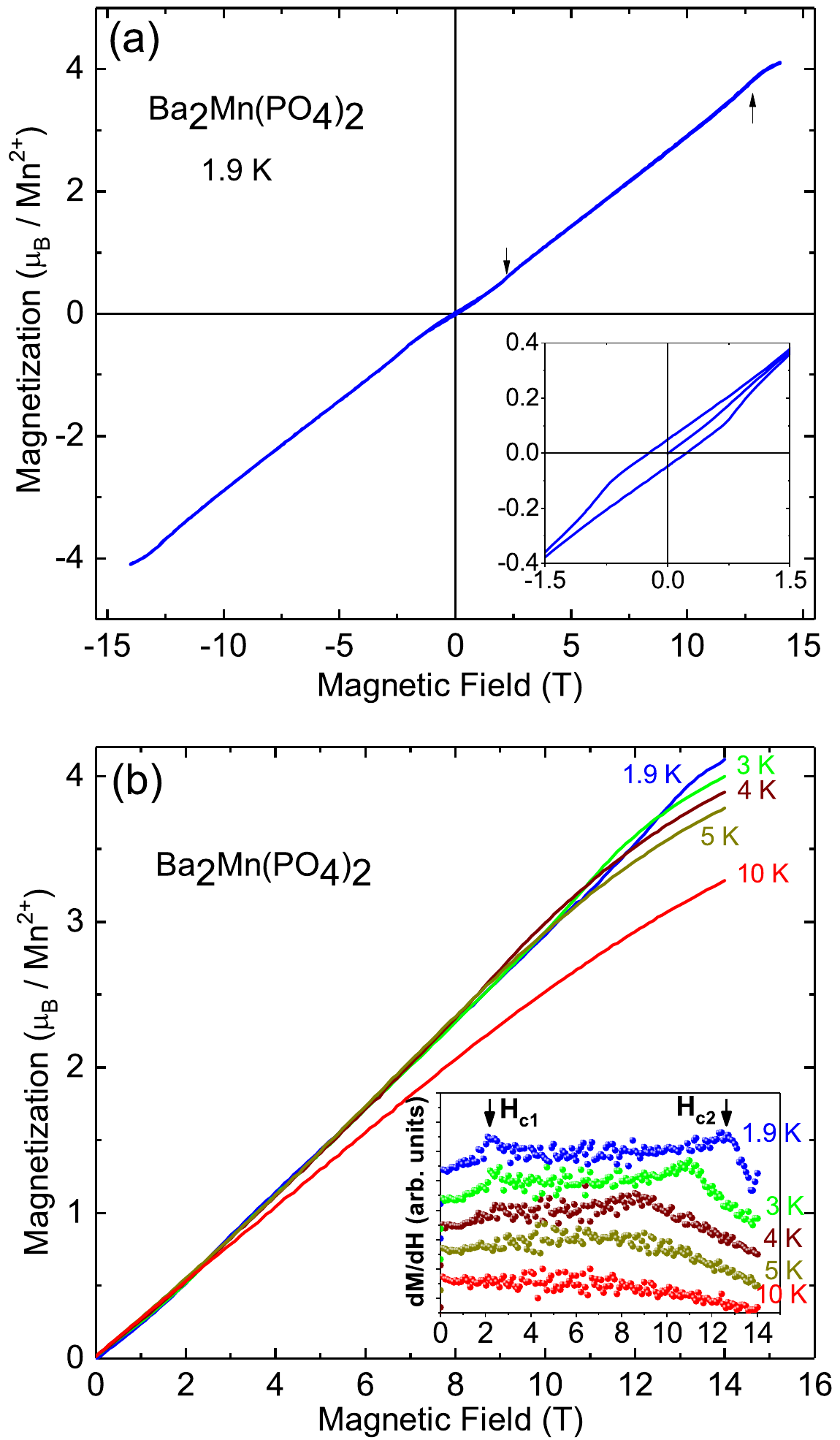}
\caption{\label{MH} (Color online) (a) Magnetization ($M$) as a function of the applied field ($H$) measured at 1.9\,K. The inset shows the zoomed \emph{M} vs.  \emph{H} curve in the low-field region. (b) The \emph{M} vs. \emph{H} curves measured at 1.9, 3, 4, 5, and 10 K. The inset shows their derivatives. The curves are vertically shifted for clarity and the arrows indicate field-induced transitions.}
\end{figure}

Isothermal magnetization curve $M(H)$ measured at 1.9\,K is shown in Fig.~\ref{MH}(a). It reveals weak hysteresis below $\sim$\,1\,T with the remanent magnetization of about 0.05\,$\mu_B$/f.u., and a change of slope at $\sim$\,2.2\,T. The small remanent magnetization is most likely due to the Mn$_3$O$_4$ impurity (see Sec.~\ref{sec:magnetization} above). At 13\,T, more than 70\,\% of the full (saturated) moment is reached, and the curve shows a small bend, which is better visible in the derivative, $dM/dH$ curve [inset of Fig.~\ref{MH} (b)].

The weak hysteretic behavior is likely due to the Mn$_3$O$_4$ impurity. The bend at 2.2\,T could be interpreted as a spin-flop transition. As for the effect at 13\,T, it may not reflect full saturation, because $M_{\rm 14 T}\sim 4.11$\,$\mu_B$/Mn$^{2+}$ yields only 82\% of the saturation value of $M_s=gS\mu_B=5$\,$\mu_B$/Mn$^{2+}$ ($S$ = $\frac{5}{2}$, $g=2$). When an anisotropic magnetization process is measured on powder sample, such bends may indicate that the system is saturated for one field direction without reaching saturation for other field directions. The single spin-flop transition at 2.2 T in \BMPO\ is in contrast to its Ni counterpart having two field induced transitions at 4 and 10 T before saturation.
%Alternatively, the 13\,T feature might indicate yet another field-induced transition, as in Ba$_2$Ni(PO$_4)_2$~\cite{Yogi2017}, and thus may require further investigation.

\subsection{Ordered states}
\subsubsection{Zero-field case}
\begin{figure}
\includegraphics[width=1.0\linewidth]{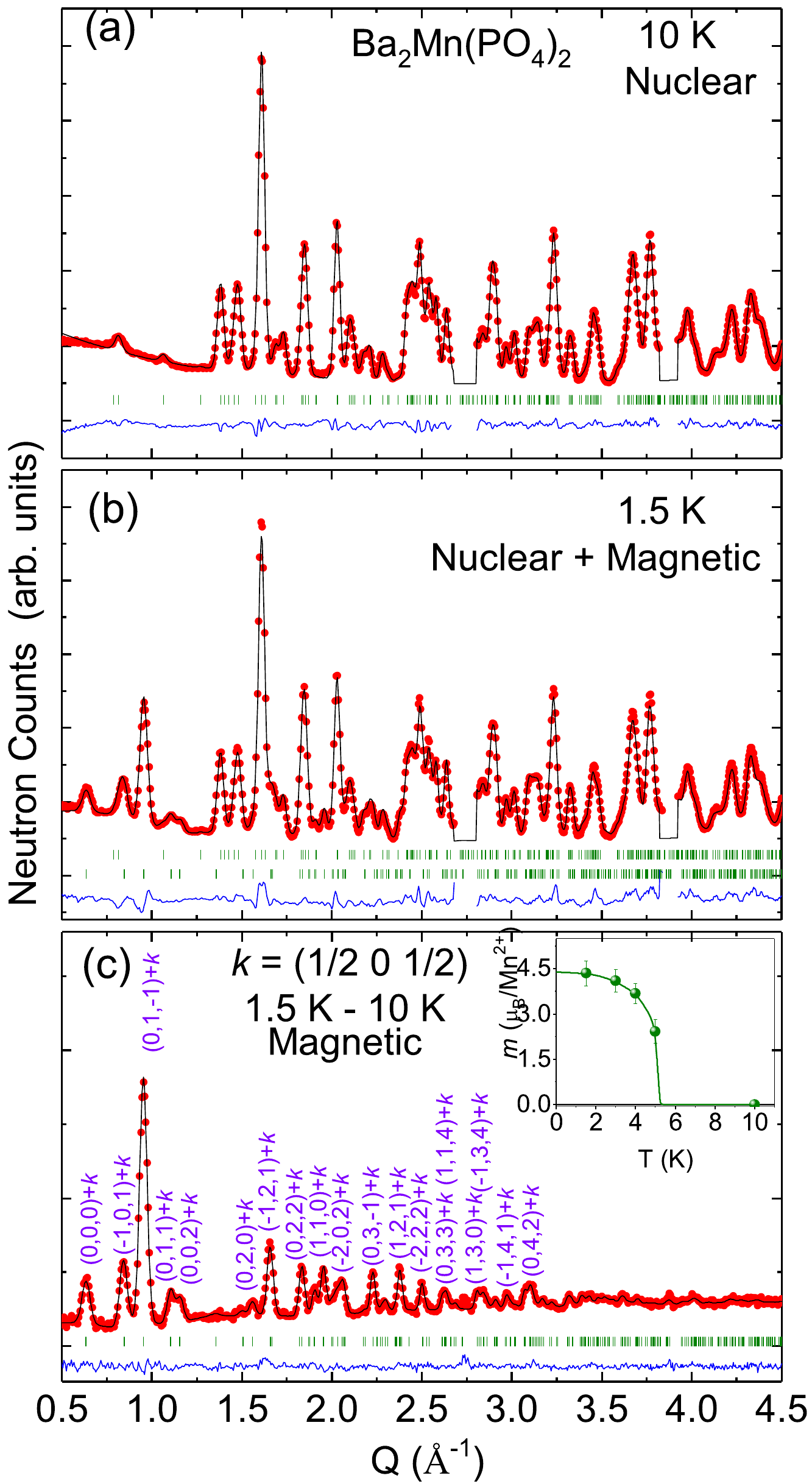}
\caption{\label{Fig:BMPO_NDP}(Color online) Experimentally observed (circles) and calculated (solid black lines) neutron diffraction patterns for \BMPO\ at (a) 10~K  (paramagnetic state) and (b) 1.5~K (magnetically ordered state), respectively measured by using an orange He cryostat. For the refinement of both the diffraction patterns, few regions are excluded where the aluminum lines (from the cryostat) were observed. (c) Magnetic pattern at 1.5~K (after subtraction of nuclear background at 10~K). The magnetic pattern in (c) is zoomed vertically by two times for clarity. The solid blue lines at the bottom of the each panel represent the difference between observed and calculated patterns. The vertical bars indicate the positions of allowed nuclear and magnetic Bragg peaks. The inset of (c) shows the temperature dependent order moment. The line is a guide to the eyes.}
\end{figure}

Neutron diffraction patterns for \BMPO\ measured at 10~K (paramagnetic state) and at 1.5~K (magnetically ordered state) are shown in Fig.~\ref{Fig:BMPO_NDP}(a) and Fig.~\ref{Fig:BMPO_NDP}(b), respectively. The refinement of the 10\,K paramagnetic pattern confirms the monoclinic crystal structure with the space group $P2_1/n$. The diffraction pattern at the lowest temperature 1.5~K (magnetic ordered state) displays a set of additional, magnetic Bragg peaks [Fig. \ref{Fig:BMPO_NDP}(b)] that can be indexed with the propagation vector ${\mathbf k}=(\frac12,0,\frac12)$ with respect to the monoclinic unit cell. This indicates a doubling of the magnetic unit cell along the $a$ and $c$ axes.
%The monoclinic crystal structure with space group $P2_1/n$ reproduce the nuclear phase at 1.5~K.

Magnetic structures of \BMPO\ compatible with the crystallographic symmetry were determined by the representation analysis using the BASIREPS program of the Fullprof suite \cite{fullprof, Bertaut.JAP.33.1138,Bertaut.ACa.24.217,Bertaut.JPC.32.C1-462,Bertaut.JMMM.24.267,Izyumov-book,Bradley_Cracknell,Cracknell,Wills.PRB.63.064430}. The symmetry analysis reveals four magnetic structures that can form upon the second-order phase transition at $T_N$. The magnetic reducible representation $\Gamma_{\rm mag}$ for the Mn site can be decomposed as a direct sum of IRs as

\begin{equation}
\label{eq:gamma_Mn}
\Gamma_{\rm mag}^{\text{Mn}}=3\Gamma_{1}^{1}+3\Gamma_{2}^{1}+3\Gamma_{3}^{1}+3\Gamma_{4}^{1}
\end{equation}

All the four $\Gamma$'s are one-dimensional and appear three times in the $\Gamma_{\rm mag}$. The basis vectors (Fourier components of the magnetization) for the magnetic Mn site $4e$ ($x$,$y$,$z$) are given in Table \ref{tab:basis_vectors_BMPO} for the four IRs. The basis vectors are calculated using the projection operator technique implemented in BASIREPS \cite{fullprof}.

\begin{table}
\caption{\label{tab:basis_vectors_BMPO} Basis vectors of the magnetic Mn site with the propagation vector ${\mathbf k}=(\frac12, 0, \frac12)$ for \BMPO. Only the real components of the basis vectors are presented. The atoms of the non-primitive basis are defined according to Mn1:~[(0.2628, 0.4911, 0.3596) : ($x$, $y$, $z$)]; Mn2:~[(0.2372, 0.9911, 0.1404) : ($\bar x+\frac12$, $y+\frac12$, $\bar z+\frac12$)]; Mn3:~[($-0.2628$, $-0.4911$, $-0.3596$): ($\bar x$, $\bar y$, $\bar z$)]; Mn4:~[(0.7628, 0.0089, 0.8596) : ($x+\frac12$, $\bar y+\frac12$, $z+\frac12$)].}
\begin{ruledtabular}
\begin{tabular}{cccccc}
%Left&Centered&Right\\
IRs &&\multicolumn{4}{c}{Basis Vectors} \\
\cline{3-6}
\\
&&Mn1 & Mn2 & Mn3 & Mn4\\
\hline
\\
$\Gamma_1^1$ & $\Psi_1$ &(100)&$(\bar 100)$&(100)&$(\bar 100)$\\
 & $\Psi_2$ &(010)&(010)&(010)&(010)\\
 & $\Psi_3$ &(001)&$(00\bar 1)$&(001)&$(00\bar 1)$\\
\\
$\Gamma_2^1$ &$\Psi_1$ &(100)&$(\bar 100)$&$(\bar 100)$&(100)\\
&$\Psi_2$&(010)&(010)&	$(0\bar 10)$&$(0\bar 10)$\\
&$\Psi_3$&(001)&(00$\bar 1)$&$(00\bar 1)$&(001)\\
\\
$\Gamma_3^1$ & $\Psi_1$ & (100)&(100)&(100)&(100) \\
&$\Psi_2$&(010)&$(0\bar 10)$&(010)&	$(0\bar 10)$\\
&$\Psi_3$&(001)&(001)&	(001)&	(001)\\
\\
$\Gamma_4^1$ & $\Psi_1$ &(100)&(100)&$(\bar 100)$&$(\bar 100)$\\
& $\Psi_2$&(010)&$(0\bar 10)$	&$(0\bar 10)$&(010)\\
& $\Psi_3$&(001)&(001)&$(00\bar 1)$&$(00\bar 1)$\\
\end{tabular}
\end{ruledtabular}
\end{table}

The refinement of the magnetic structure was tested for all the four $\Gamma$'s. Only the $\Gamma_1$ produced a good fit of the observed diffraction patterns at 1.5 K. The fitted pattern is shown in Fig. \ref{Fig:BMPO_NDP}(b). For further clarification, the pure magnetic pattern at 1.5~K (after subtraction of the nuclear background at 10~K) is shown in Fig. \ref{Fig:BMPO_NDP}(c) along with the calculated magnetic pattern. The $R_{mag}$ factor was found to be 6.92~$\%$.

\begin{figure}
\includegraphics[width=1\linewidth]{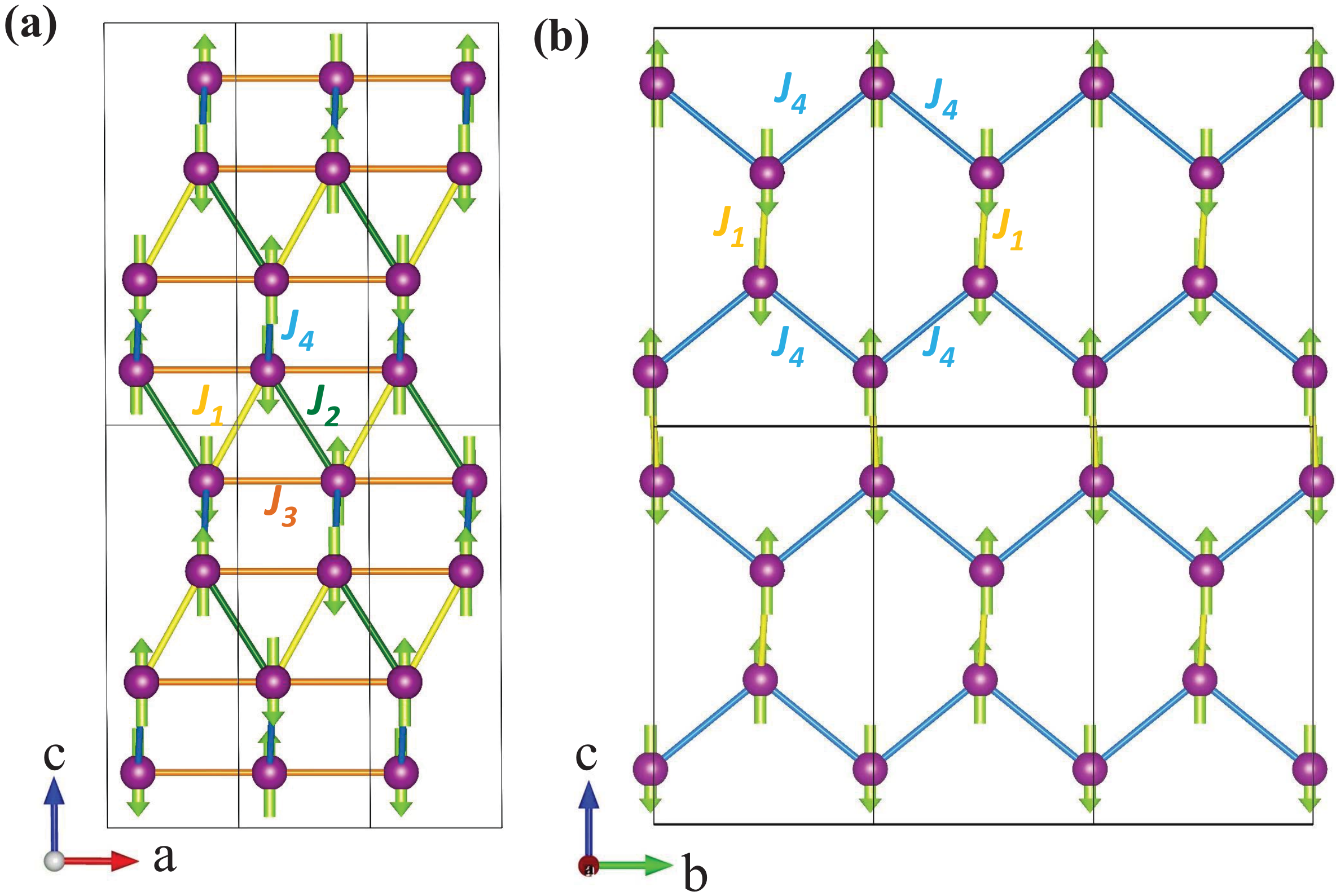}
\caption{\label{Fig:mag_struc_BMPO}(Color online) The magnetic structure of \BMPO. The projection of the magnetic structure in the (a) $ac$ and (b) $bc$ planes, respectively.}
 \end{figure}

The basis vectors for $\Gamma_1$ [Table~\ref{tab:basis_vectors_BMPO}] indicate that all three components of the magnetic moment can be refined. However, we found that the $c$-component alone is sufficient to fit the data. The refinement of the $a$- and $b$-components did not improve the fit, while their refined values remained close to zero. We thus conclude that magnetic moments point along the $c$-direction of the crystal structure. The ordered moment is $4.33\pm0.08$\,$\mu_B$/Mn$^{2+}$ at 1.5\,K, slightly lower than 5\,$\mu_B$/Mn$^{2+}$ expected for $S=\frac52$. This reduction may reflect residual fluctuations in the ordered state due to geometrical spin frustration or the Mn--O hybridization that reduces the ordered moment on Mn$^{2+}$.

The resulting magnetic structure is shown in Fig. \ref{Fig:mag_struc_BMPO}. Within the zigzag chains, two nearest-neighbor spins align parallel on the $J_1$ bond and antiparallel on the $J_2$ bond. This leads to an antiparallel alignment of the spins on the $J_3$ bond, suggesting that $J_2$ and $J_3$ dominate over $J_1$ (all couplings are antiferromagnetic, see Sec.~\ref{sec:dft} below). Regarding the honeycomb structural units, this magnetic order can be viewed as stripe or anti-N\'{e}el type, similar to the ground-state spin configuration in \BNPO~\cite{Yogi2017,note1}. The magnetic moments are fully compensated, and zero net magnetization is expected, in agreement with our DFT results reported in Sec.~\ref{sec:dft} below.

\subsubsection{Field-induced states}
\begin{figure*}
\includegraphics[width=0.80\linewidth]{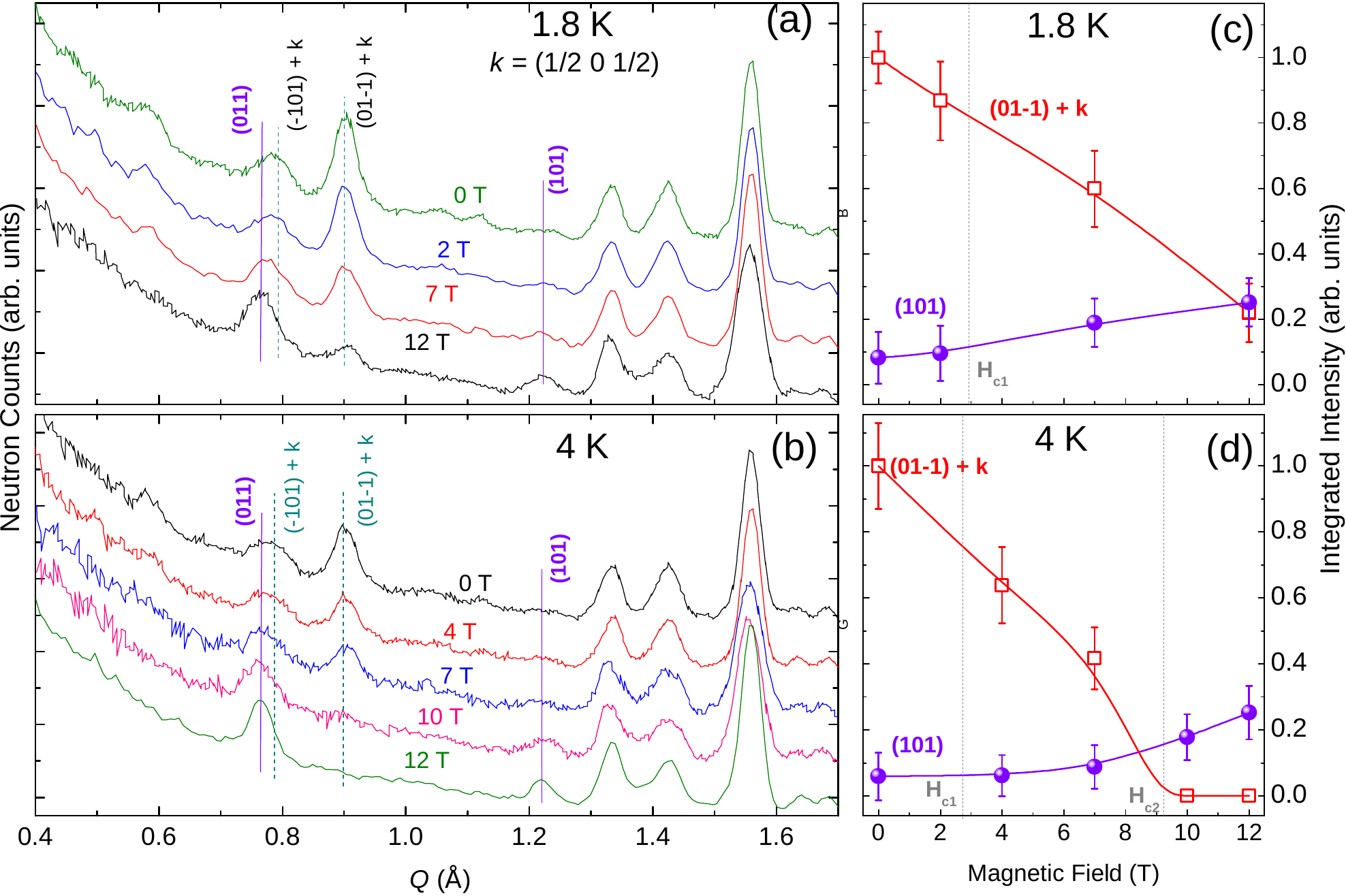}
\caption{\label{Fig:patterns_Magfiled} (Color online) Field evolution of the neutron diffraction patterns at (a) 1.8\,K and (b) 4\,K. The field-dependent integrated intensity of the magnetic peaks $(01\bar 1)+{\mathbf k}$ and (011) at (c) 1.8\,K and (d) 4\,K.}
\end{figure*}

To explore the field-induced states, we performed neutron diffraction measurements under several magnetic fields at 1.8 and 4~K (Fig.~\ref{Fig:patterns_Magfiled}). With increasing magnetic field, the intensity of the magnetic peaks characteristic of the AFM phase decreases monotonically and at 12\,T becomes very small at 1.8\,K, whereas at 4\,K it vanishes. On the other hand, additional magnetic intensity appears on top of some of the nuclear Bragg peaks indicating the formation of a ferromagnetic phase. Field dependence of the characteristic magnetic peaks $(01\bar 1)+{\mathbf k}$ and (101) at 1.8 and 4~K is shown in panels (c) and (d) of Fig.~\ref{Fig:patterns_Magfiled}, respectively.

\begin{figure}
\includegraphics[width=1.00\linewidth]{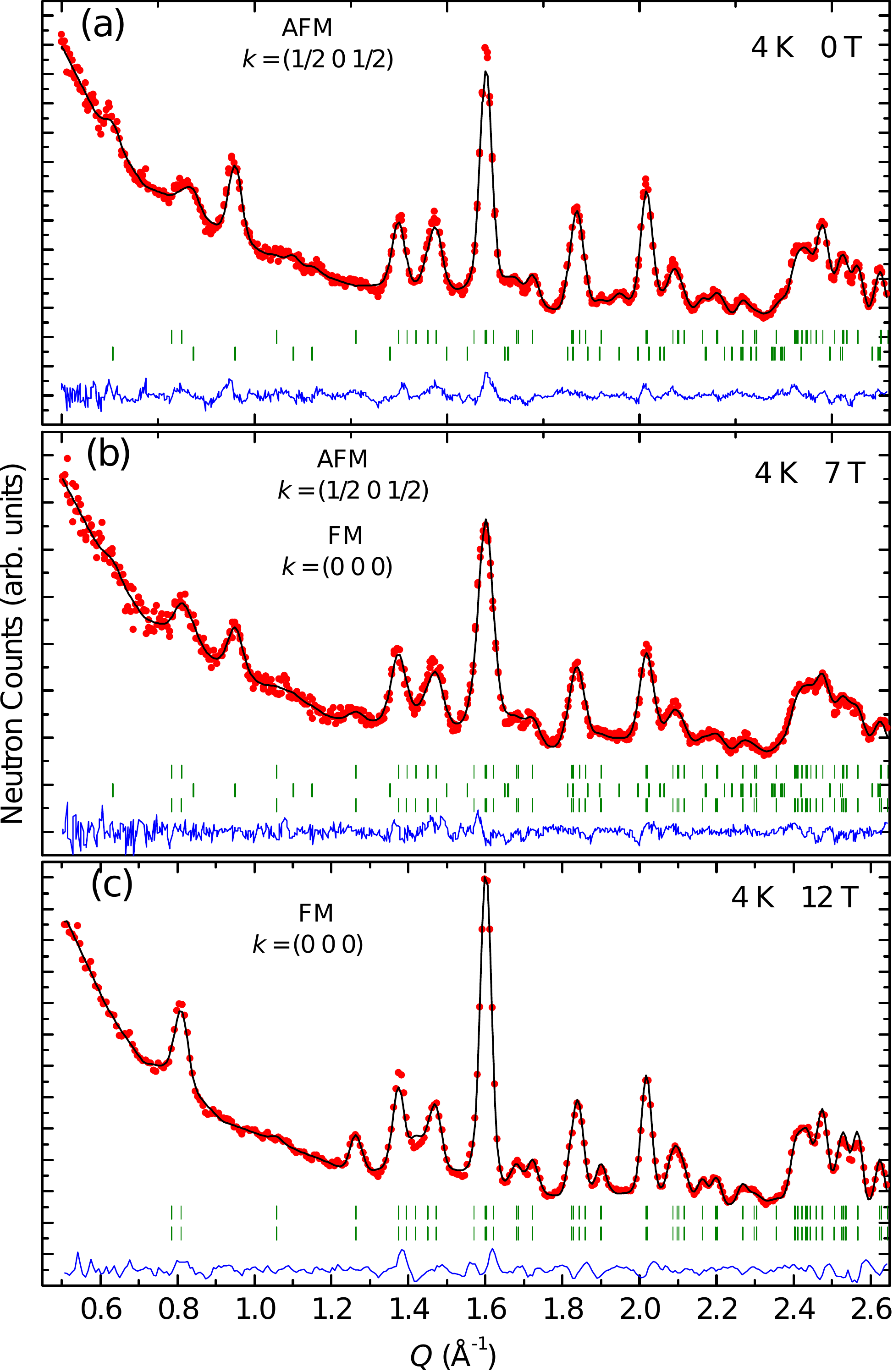}
\caption{\label{Fig:fitted_patterns_Magfiled} (Color online) Experimentally observed (circles) and calculated (solid black lines) neutron diffraction patterns for \BMPO\ at 4~K under an applied magnetic field of (a) 0~T  (pure AFM state), (b) 7~T (AFM+FM state) and  (c) 12~T (pure FM state), respectively. The solid blue lines at the bottom of the each panel represent the difference between observed and calculated patterns. The vertical bars indicate the positions of allowed nuclear and magnetic Bragg peaks. }
\end{figure}

\begin{figure}
\includegraphics[width=1.00\linewidth]{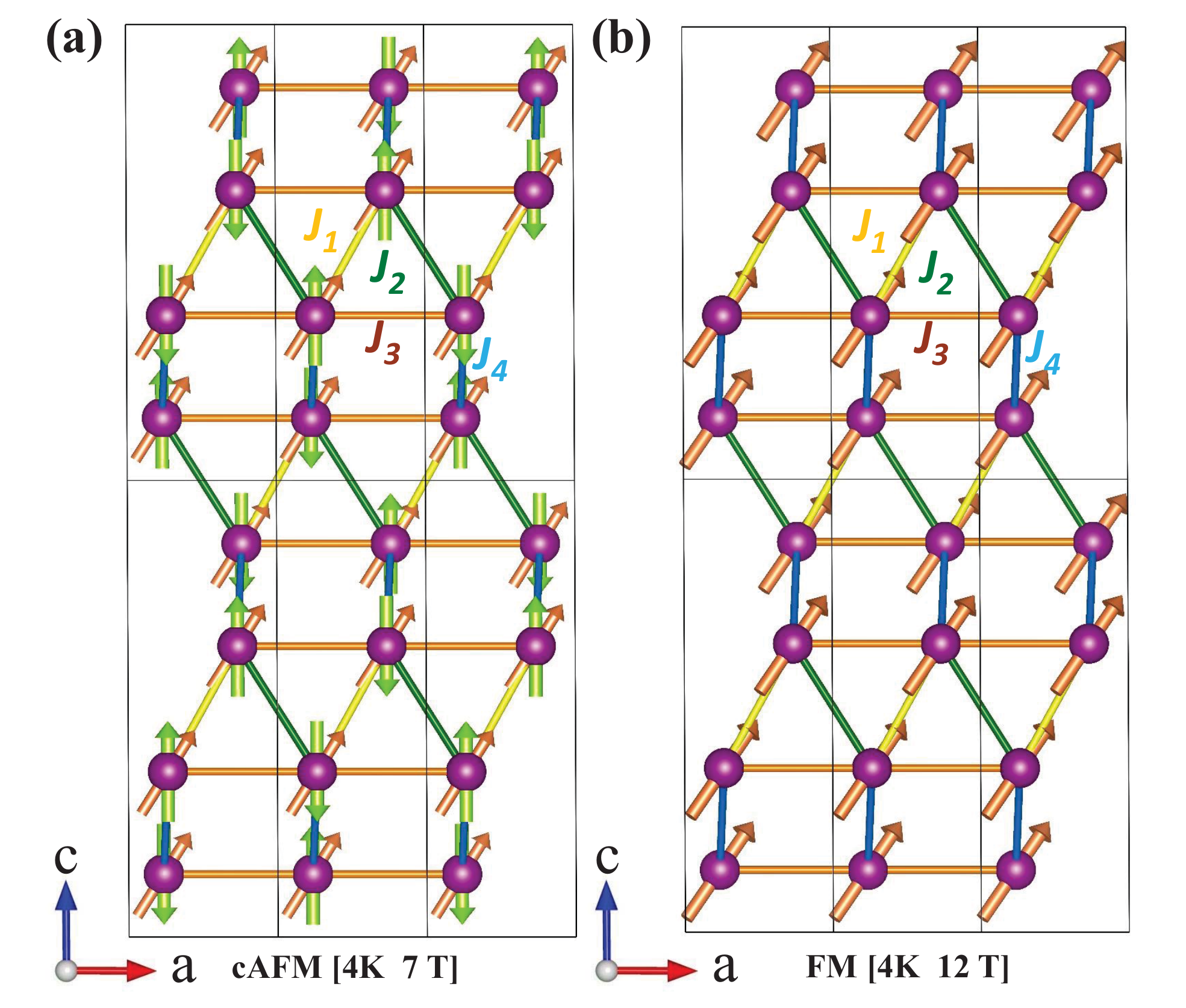}
\caption{\label{Fig:MAgstruc_FM} (Color online) Schematic representation of the magnetic structure corresponding to field-induced (a) canted AFM state and (b) polarized FM state in $ac$-plane, respectively.}
\end{figure}

The Rietveld-refined neutron diffraction patterns measured under 0, 7 and 12~T at 4\,K are shown in Fig.~\ref{Fig:fitted_patterns_Magfiled}. The pattern in zero field could be described by a superposition of the nuclear phase and AFM phase with ${\mathbf k}=(\frac 12, 0, \frac12)$]. In contrast, under the field of 12\,T, only magnetic peaks corresponding to the propagation vector [${\mathbf k}$ = (0 0 0)] are present, indicating that a field-induced ferromagnetic phase has formed. Good agreement with the experimental pattern is obtained by a refinement with a two-phase model (nuclear and FM phases). The refined magnetic moments are found to be $m_a = -2.4\pm0.3~\mu_B$, $m_b = 2.4\pm0.3~\mu_B$, and $m_c = 3.5\pm0.2~\mu_B$ with the total moment of $m_{\rm Mn} = 4.9$\,$\mu_B$, which is very close to the spin-only value of 5\,$\mu_B$/Mn$^{2+}$. Schematic representation of the magnetic structure in this phase is given in Fig. \ref{Fig:MAgstruc_FM} (b).

In the field-induced intermediate state (under 7~T), magnetic intensities are distributed between the ${\mathbf k}=(\frac 12
, 0, \frac12)$] and FM [${\mathbf k}$ = (0 0 0)] peaks. A three-phase refinement with the nuclear, AFM and FM phases results in a good description of the experimental data. During the refinement, same scale factor was considered for all the three phases. A schematic representation of the magnetic structure for this intermediate field range is shown in Fig.~\ref{Fig:MAgstruc_FM} (a). The coexistence of the AFM and FM phases simply represents a canted state characterized by two propagation vectors.

\subsection{Magnetic phase diagram}
\label{sec:phase-diagram}
\begin{figure}
\includegraphics[width=1.0\linewidth]{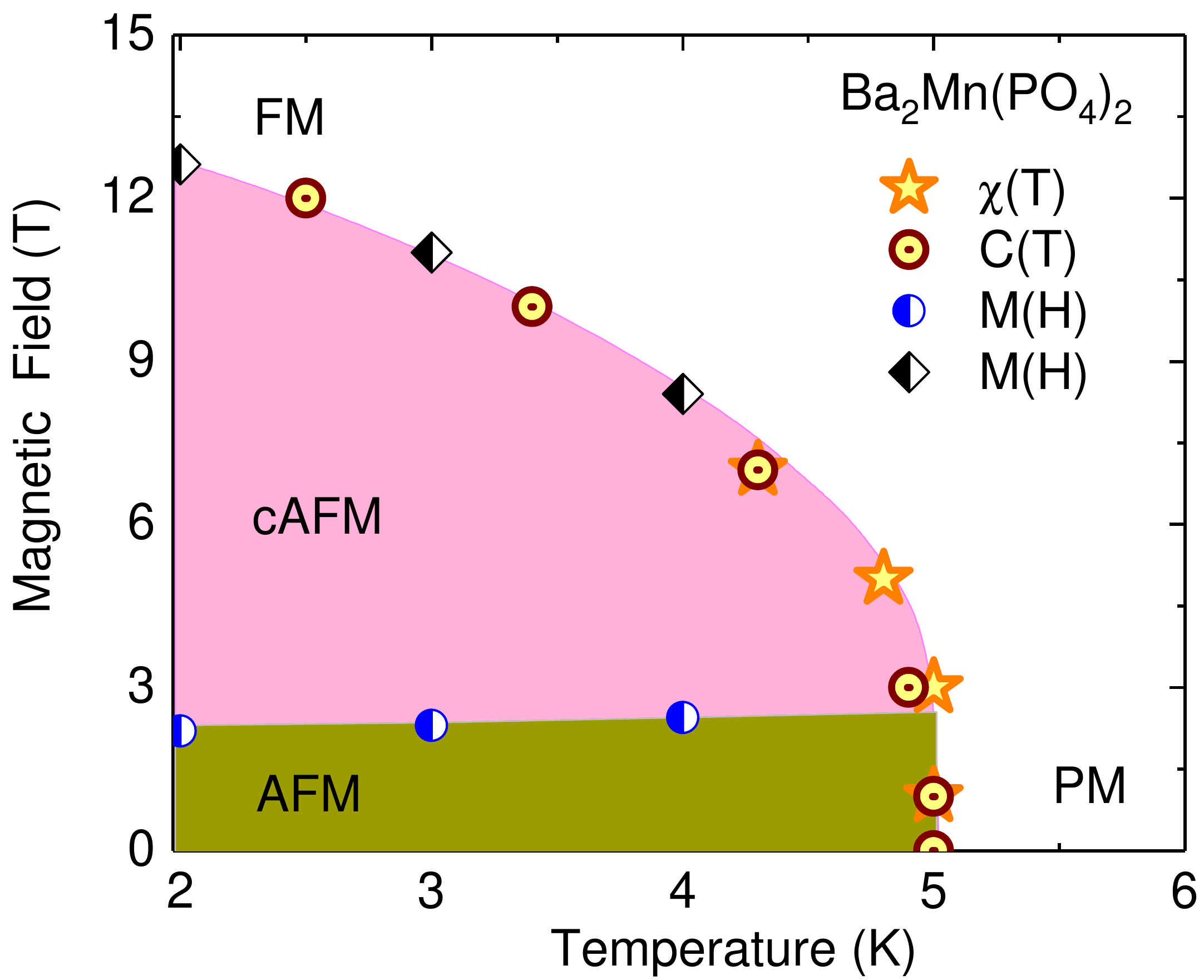}
\caption{\label{Fig:phase-diagram} (Color online) Magnetic phase diagram of Ba$_2$Ni(PO$_4)_2$ in the $H-T$ plane. The half-filled and open circles with central dot points are obtained from magnetization and heat-capacity measurements, respectively. The points represented by stars are obtained from the susceptibility ($\chi$) curves.}
\end{figure}

The magnetic phase diagrams in the \emph{H}--\emph{T} plane for \BMPO\ is shown in Fig. \ref{Fig:phase-diagram}. In zero field, a magnetic phase transition occurs from paramagnetic state to the ordered AFM ground state at $\sim$ 5~K. With increasing magnetic field, the transition temperature remains constant over the low-field region (below $\sim$ 3~T), then decreases rapidly, and drops below 2~K above $\sim$ 12~T. In the ordered state (below 5 K), with increasing magnetic field, a spin-flop transition from the AFM state to a field induced state is found at $\sim$ 2.2 T. The ordered state at intermediate fields features canted spins that gradually tilt toward the field direction. Neutron powder diffraction sees this state as a combination of AFM and FM phases with different propagation vectors. All the temperature and field induced transitions are of second order as no hysteresis is evident across the phase transitions. \\

\subsection{Microscopic magnetic model}
\label{sec:dft}
(Color online) Magnetic exchange couplings calculated for \BMPO\ are summarized in Table~\ref{tab:exchange}. They are AFM and compatible with the experimental Curie-Weiss temperature $\theta_{\rm CW}\simeq -12$\,K estimated as
\begin{equation}
 \theta_{\rm CW}=-\frac{S(S+1)}{3}\sum_i z_iJ_i,
\end{equation}
where $z_i$ is the number of couplings per site. Similar to \BNPO, the leading coupling is $J_3$, whereas $J_1$ and $J_2$ are of similar size and frustrated. The $J_2>J_1$ condition can be reproduced in different band-structure codes (Table~\ref{tab:exchange}). It is also robust with respect to the variation of the Coulomb repulsion parameter $U_d$ within the realistic $4-6$\,eV range. We thus conclude that $J_2$ and $J_3$ overcome $J_1$ resulting in parallel spin alignment on the $J_1$ bonds despite the AFM nature of this coupling.

\begin{table}
\caption{\label{tab:exchange}
Exchange couplings $J_i$ (in\,K) and the ensuing Curie-Weiss temperature $\theta_{\rm CW}$ (in\,K) calculated using the \texttt{FPLO} and \texttt{VASP} codes. The Mn--Mn distances $d_{\rm Mn-Mn}$ are given in\,\r A.
}
\begin{ruledtabular}
\begin{tabular}{cccc}
      & $d_{\rm Mn-Mn}$ & $J_i$ (\texttt{FPLO}) & $J_i$ (\texttt{VASP}) \\
$J_1$ & 5.089 & 0.5 & 0.7 \\
$J_2$ & 5.286 & 0.8 & 1.0 \\
$J_3$ & 5.311 & 1.6 & 1.7 \\
$J_4$ & 5.807 & 0.3 & 0.3 \\
$\theta_{\rm CW}$ & & $-14.9$ & $-16.6$ \\
\end{tabular}
\end{ruledtabular}
\end{table}

\begin{figure}
\includegraphics{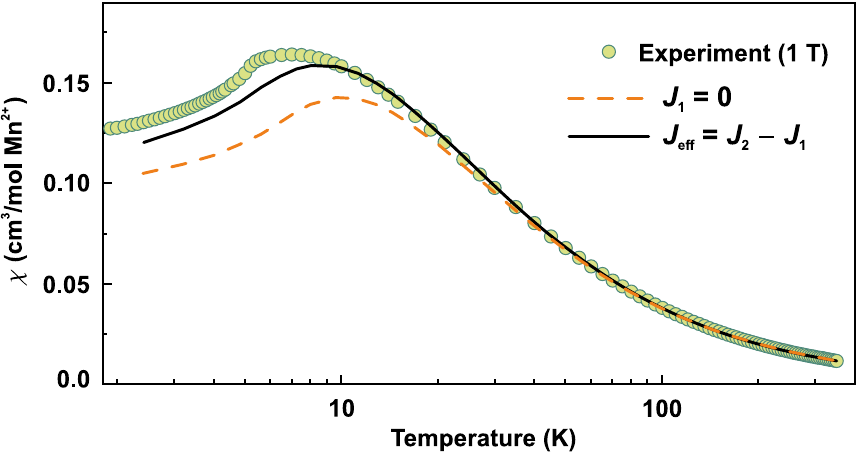}
\caption{\label{fig:fit}
Magnetic susceptibility of \BMPO\ measured in the applied field of 1\,T, and its fit with two models: i) without frustration ($J_1=0$, dashed line); ii) with frustration taken into account on the mean-field level via the effective coupling $J_{\rm eff}=J_2-J_1$ replacing $J_2$ (solid line), see text for details.}
\end{figure}

The competition between $J_1$ and $J_2$ has strong impact on thermodynamic properties of \BMPO. We first simulated magnetic susceptibility of \BMPO\ disregarding the frustrating coupling $J_1$ completely. Despite satisfactory agreement with the experimental data at high temperatures, deviations occur already at 30\,K (Fig.~\ref{fig:fit}, dashed line), whereas the calculated $T_N\simeq 7.5$\,K is well above the experimental value. On the mean-field level, the effect of frustration is merely the renormalization of $J_2$ to $J_{\rm eff}=J_2-J_1\simeq 0.3$\,K. With this effective coupling, we can largely improve the fit of the magnetic susceptibility (Fig.~\ref{fig:fit}, solid line)~\cite{note2} and obtain $T_N\simeq 5.5$\,K in nearly perfect agreement with 5.0\,K determined experimentally.

%D=0.285 K --> 45.6 mK
%E=0.105 K --> 16.8 mK

Turning now to the anisotropic part of the spin Hamiltonian, we first compare energies for different spin directions on a single Mn$^{2+}$ ion. The $c$ direction is preferable, whereas spins directed along $a$ and $b$ increase the energy by 0.39\,K and 0.18\,K, respectively. This corresponds to the single-ion terms $D=45.6$\,mK and $E=16.8$\,mK, where $z=c$, $x=a$, and $y=b$ in the spin Hamiltonian, Eq.~\eqref{eq:hamiltonian}. For comparison, we considered lowest-energy AFM configurations with all spins directed along $c$ or $a$, and found the energy difference of 0.41\,K/Mn$^{2+}$, which includes both single-ion effects and the exchange anisotropy (different $J_{ij}$'s for different spin components). The single-ion anisotropy turns out to be the leading term that chooses $c$ as preferred spin direction, in agreement with the direction of the magnetic moment determined experimentally. Exchange anisotropy is only 6\,\% of the single-ion term.

Finally, we inspected possible spin canting in the AFM ground state. To this end, we performed a calculation without constraining spin directions and arrived at $m_c=4.64$\,$\mu_B$ as well as $m_b<0.001$\,$\mu_B$. This confirms the absence of any appreciable spin canting in \BMPO, whereas small remanent magnetization detected in the magnetization measurements should be due to the Mn$_3$O$_4$ impurity.

\section{Discussion and Summary}
\BMPO\ and \BNPO\ are isostructural. They share the same type of AFM order and even feature similar N\'eel temperatures $T_N\simeq 5$\,K. Nevertheless, their microscopic scenarios are distinctly different. Isotropic exchange interactions $J_i$ follow the same pattern in both compounds, with the leading coupling $J_3$ and a sizable ``diagonal'' couplings $J_1$ and $J_2$. Our DFT based calculations for \BMPO\ reveal that all the exchange interactions ($J_1$-$J_4$) are AFM in nature (Table~\ref{tab:exchange}).

As per the Table~\ref{tab:exchange}, the interconnection between Mn-Mn bonds forms a frustrated zigzag spin chain lattice made up by three different exchange couplings ($J_1$-$J_3$). Further, the pathways of the exchange couplings are composed of super-exchange interaction by two different P(1)O$_{4}$ and P(2)O$_{4}$ tetrahedra, respectively. The possible superexchange interaction pathways in \BMPO\ are $J_1$  [Mn-O1-P(1)-O2-Mn], $J_2$ [Mn-O1-P(1)-O4-Mn], $J_3$ [Mn-O2/O3-P(1)-O4-Mn] and $J_4$ [Mn-O6-P(2)-O8-Mn] and the distance of respective exchange couplings are listed in the Table~\ref{tab:exchange}. Although they are having similar super exchange pathways, their strength are quite different as estimated by DFT calculations.

By considering the hierarchy of the exchange interactions, the magnetic lattice of \BMPO\ can be defined as a frustrated zigzag chain along the $a$-axis. The strongest exchange interaction $J_3$ resulted into an antiparallel alignment of the second nearest-neighbour spins along the zigzag chain. The spin on the nearest neighbour site, that forms a triangular lattice with the antiparallelly aligned two spins, becomes now frustrated as both the $J_1$ and $J_2$ are antiferromagnetic. The relatively stronger $J_2$ overcomes $J_1$ and resulting into the observed magnetic ground state where the nearest neighbour spin has antiparallel alignment with respect to the $J_2$ and parallel alignment with respect to the $J_1$ within the zigzag chain.

In \BNPO, $J_1$ is a small coupling of minor importance that is easily overridden by $J_2$. On the other hand, single-ion anisotropy turns out to be as strong as the isotropic couplings, and competition between single-ion anisotropies on different Ni$^{2+}$ sites gives rise to a rather unusual frustrated scenario that manifests itself in the short-range order above $T_N$.

\BMPO\ behaves more like a conventional antiferromagnet without any short-range order above $T_N$. Its single-ion anisotropy is less than 5\,\% of $J_3$ and appears to be uniaxial, with same preferred direction for all Mn$^{2+}$ ions. Although lacking frustration related to the anisotropy, \BMPO\ exhibits appreciable frustration by isotropic exchange couplings, because $J_1$ and $J_2$ are now of similar size. This frustration has a strong impact on thermodynamic properties and $T_N$.

Despite different mechanisms of the frustration, its magnitude turns out to be similar in \BMPO\ and \BNPO, with the comparable $|\theta_{\rm CW}$/$T_N|$~=~2.4 and 2.1, respectively. The manifestations are different, though. \BNPO\ with its frustration due to magnetic anisotropy exhibits a short-range magnetic order above $T_N$, whereas \BMPO\ with its frustration by isotropic exchange couplings lacks such a short-range order. Moreover, \BNPO\ undergoes two consecutive magnetic transitions in zero field, whereas \BMPO\ shows a single transition.

Field-induced behavior bears more dissimilarities. The horizontal line on the $H-T$ phase diagram at 2.5\,T (\BMPO) and 4\,T (\BNPO) would be consistent with a spin-flop transition. In \BMPO, we indeed observe a gradual spin canting from the AFM state toward the fully polarized state, without any peculiarities, whereas in \BNPO\ a distinct high-field phase above 10\,T was reported~\cite{Yogi2017}. For \BMPO, the behavior in higher fields indicates full polarization. Our neutron data for \BMPO\ at 12\,T and 4\,K are compatible with the simple ferromagnetic state. On the other hand, magnetization does not show full polarization under 14\,T at 1.9\,K. This discrepancy most likely reflects directional dependence of the saturation field.

In summary, we performed a comprehensive characterization of the frustrated spin-$\frac52$ antiferromagnet \BMPO. Its AFM order is driven by couplings between the frustrated 1D zigzag spin chain. The anisotropy is mostly of single-ion nature and weak. On the other hand, frustration of isotropic couplings within and between the 1D zigzag spin chain has a strong impact on thermodynamic properties. We conclude that different mechanisms of the frustration, competing isotropic exchange couplings or mismatched single-ion anisotropies, can lead to very different physical scenarios. The Mn and Ni compounds represent two limiting cases, whereas the isostructural Co compound~\cite{Faza2001} may combine two types of frustration and will be interesting for future studies.

\acknowledgments
AY would like to acknowledge TIFR, Mumbai for the financial support. AT would like to acknowledge DAE, India for financial support. AAT was supported by the Federal Ministry for Education and Research through the Sofja Kovalevskaya Award of Alexander von Humboldt Foundation. Fig.~\ref{fig:structure} and Fig.~\ref{Fig:mag_struc_BMPO} were prepared using the \texttt{VESTA} software~\cite{vesta}.


\begin{thebibliography}{80}%
\makeatletter
\providecommand \@ifxundefined [1]{%
 \@ifx{#1\undefined}
}%
\providecommand \@ifnum [1]{%
 \ifnum #1\expandafter \@firstoftwo
 \else \expandafter \@secondoftwo
 \fi
}%
\providecommand \@ifx [1]{%
 \ifx #1\expandafter \@firstoftwo
 \else \expandafter \@secondoftwo
 \fi
}%
\providecommand \natexlab [1]{#1}%
\providecommand \enquote  [1]{``#1''}%
\providecommand \bibnamefont  [1]{#1}%
\providecommand \bibfnamefont [1]{#1}%
\providecommand \citenamefont [1]{#1}%
\providecommand \href@noop [0]{\@secondoftwo}%
\providecommand \href [0]{\begingroup \@sanitize@url \@href}%
\providecommand \@href[1]{\@@startlink{#1}\@@href}%
\providecommand \@@href[1]{\endgroup#1\@@endlink}%
\providecommand \@sanitize@url [0]{\catcode `\\12\catcode `\$12\catcode
  `\&12\catcode `\#12\catcode `\^12\catcode `\_12\catcode `\%12\relax}%
\providecommand \@@startlink[1]{}%
\providecommand \@@endlink[0]{}%
\providecommand \url  [0]{\begingroup\@sanitize@url \@url }%
\providecommand \@url [1]{\endgroup\@href {#1}{\urlprefix }}%
\providecommand \urlprefix  [0]{URL }%
\providecommand \Eprint [0]{\href }%
\providecommand \doibase [0]{http://dx.doi.org/}%
\providecommand \selectlanguage [0]{\@gobble}%
\providecommand \bibinfo  [0]{\@secondoftwo}%
\providecommand \bibfield  [0]{\@secondoftwo}%
\providecommand \translation [1]{[#1]}%
\providecommand \BibitemOpen [0]{}%
\providecommand \bibitemStop [0]{}%
\providecommand \bibitemNoStop [0]{.\EOS\space}%
\providecommand \EOS [0]{\spacefactor3000\relax}%
\providecommand \BibitemShut  [1]{\csname bibitem#1\endcsname}%
\let\auto@bib@innerbib\@empty
%</preamble>

\bibitem{Kanamori1959} J. Kanamori, Superexchange interaction and symmetry properties of electron orbitals, J. Phys. Chem. Solids \textbf{10}, 87 (1959).

\bibitem{Anderson1959} P.W. Anderson, New Approach to the Theory of Superexchange Interactions, Phys. Rev. \textbf{115}, 2 (1959).

\bibitem{Goodenough} J.B. Goodenough, \emph{Magnetism and the Chemical Bond} (Interscience Publishers, New York, 1963).

\bibitem{Ranjith2015} K.M. Ranjith, R. Nath, M. Skoulatos, L. Keller, D. Kasinathan, Y. Skourski, and A.A. Tsirlin, Collinear order in the frustrated three-dimensional $\text{spin}\ensuremath{-}\frac{1}{2}$ antiferromagnet ${\mathrm{Li}}_{2}{\mathrm{CuW}}_{2}{\mathrm{O}}_{8}$, Phys. Rev. B \textbf{92}, 094426 (2015).

\bibitem{Ranjith2016} K.M. Ranjith, R. Nath, M. Majumder, D. Kasinathan, M. Skoulatos, L. Keller, Y. Skourski, M. Baenitz, and A.A. Tsirlin, Commensurate and incommensurate magnetic order in spin-1 chains stacked on the triangular lattice in ${\mathrm{Li}}_{2}{\mathrm{NiW}}_{2}{\mathrm{O}}_{8}$, Phys. Rev. B \textbf{94}, 014415 (2016).

\bibitem{Nath2014} R. Nath, K. M. Ranjith, B. Roy, D. C. Johnston, Y. Furukawa, and A. A. Tsirlin, Magnetic transitions in the spin-5/2 frustrated magnet BiMn$_2$PO$_6$ and strong lattice softening in BiMn$_2$PO$_6$ and BiZn$_2$PO$_6$ below 200 K, Phys. Rev. B \textbf{90}, 024431 (2014).

\bibitem{Mentre2009} O. Mentr\'e, E. Janod, P. Rabu, M. Hennion, F. Leclercq-Huguex, J. Kang, C. Lee, M.-H. Whangbo, and S. Petit, Incommensurate spin correlation driven by frustration in ${\text{BiCu}}_{2}{\text{PO}}_{6}$, Phys. Rev. B \textbf{80}, 180413(R) (2009).

\bibitem{Koteswararao2007} B. Koteswararao, S. Salunke, A.V. Mahajan, I. Dasgupta, and J. Bobroff, Spin-gap behavior in the two-leg spin-ladder ${\text{BiCu}}_{2}{\text{PO}}_{6}$, Phys. Rev. B \textbf{76}, 052402 (2007).

\bibitem{Tsirlin2010} A.A. Tsirlin, I. Rousochatzakis, D. Kasinathan, O. Janson, R. Nath, F. Weickert, C. Geibel, A.M. L\"auchli, and H. Rosner, Bridging frustrated-spin-chain and spin-ladder physics: Quasi-one-dimensional magnetism of ${\text{BiCu}}_{2}{\text{PO}}_{6}$, Phys. Rev. B \textbf{82}, 144426 (2010).

\bibitem{Plumb2013} K. Plumb, Z. Yamani, M. Matsuda, G.J. Shu, B. Koteswararao, F.C. Chou, and Y.-J. Kim, Incommensurate dynamic correlations in the quasi-two-dimensional spin liquid ${\text{BiCu}}_{2}{\text{PO}}_{6}$, Phys. Rev. B \textbf{88}, 024402 (2013).

\bibitem{Plumb2016} K.W. Plumb, K. Hwang, Y. Qiu, L.W. Harriger, G.E. Granroth, A.I. Kolesnikov, G.J. Shu, F.C. Chou, Ch. R\"uegg, Y.B. Kim, and Y.-J. Kim, Quasiparticle-continuum level repulsion in a quantum magnet, Nature Phys. \textbf{12}, 224 (2016).

\bibitem{Kohama2012} Y. Kohama, S. Wang, A. Uchida, K. Prsa, S. Zvyagin, Y. Skourski, R.D. McDonald, L. Balicas, H.M. Ronnow, C. R\"uegg, and M. Jaime, Anisotropic Cascade of Field-Induced Phase Transitions in the Frustrated Spin-Ladder System ${\text{BiCu}}_{2}{\text{PO}}_{6}$, Phys. Rev. Lett. \textbf{109}, 167204 (2012).

\bibitem{Kohama2014} Y. Kohama, K. Mochidzuki, T. Terashima, A. Miyata, A. DeMuer, T. Klein, C. Marcenat, Z.L. Dun, H. Zhou, G. Li, L. Balicas, N. Abe, Y.H. Matsuda, S. Takeyama, A. Matsuo, and K. Kindo, Entropy of the quantum soliton lattice and multiple magnetization steps in ${\text{BiCu}}_{2}{\text{PO}}_{6}$, Phys. Rev. B \textbf{90}, 060408(R) (2014).

\bibitem{Yogi2017} A. Yogi, A. K. Bera, A. Maurya, Ruta Kulkarni, S. M. Yusuf, A. Hoser, A. A. Tsirlin, and A. Thamizhavel, Stripe order on the spin-1 stacked honeycomb lattice in Ba$_{2}$Ni(PO$_{4}$)$_{2}$, Phys. Rev. B \textbf{95}, 024401 (2017).

\bibitem{fullprof} J. Rodriguez-Carvajal, Recent advances in magnetic structure determination by neutron powder diffraction, Physica B: Condensed Matter \textbf{192}, 55 (1993).

\bibitem{fplo} K. Koepernik, H. Eschrig, Full-potential nonorthogonal local-orbital minimum-basis band-structure scheme, Phys. Rev. B \textbf{59}, 1743 (1999).

\bibitem{vasp1} G. Kresse and J. Furthm\"uller, Efficiency of ab-initio total energy calculations for metals and semiconductors using a plane-wave basis set, Computational Materials Science \textbf{6}, 15 (1996).

\bibitem{vasp2} G. Kresse and J. Furthm\"uller, Efficient iterative schemes for \textit{ab initio} total-energy calculations using a plane-wave basis set, Phys. Rev. B \textbf{54}, 11169 (1996).

\bibitem{pbe96} J.P. Perdew, K. Burke, and M. Ernzerhof, Generalized Gradient Approximation Made Simple, Phys. Rev. Lett. \textbf{77}, 3865 (1996).

\bibitem{Xiang2013} H. Xiang, C. Lee, H.-J. Koo, X. Gong, and M.-H. Whangbo, Magnetic properties and energy-mapping analysis, Dalton. Trans. \textbf{42}, 823 (2013).

\bibitem{loop} S. Todo and K. Kato, Cluster algorithms for general-$S$ quantum spin systems, Phys. Rev. Lett. \textbf{87}, 047203 (2001).

\bibitem{alps} A.F. Albuquerque, F. Alet, P. Corboz, P. Dayal, A. Feiguin, S. Fuchs, L. Gamper, E. Gull, S. G\"urtler, A. Honecker, R. Igarashi, M. K\"orner, A. Kozhevnikov, A. L\"auchli, S.R. Manmana, M. Matsumoto, I.P. McCulloch, F. Michel, R.M. Noack, G. Paw{\l}owski, L. Pollet, T. Pruschke, U. Schollw\"ock, S. Todo, S. Trebst, M. Troyer, P. Werner, and S. Wessel, The ALPS project release 1.3: Open-source software for strongly correlated systems, J. Magn. Magn. Mater. \textbf{310}, 1187 (2007).

\bibitem{Faza2001} N. Faza, W. Treutmann, and D. Babel, Structural and Magnetochemical Studies at the Ternary Phosphates Ba$_{2}$M$^{II}$(PO$_{4}$)$_{2}$ (M$^{II}$ = Mn, Co) and Refinement of the Crystal Structure of BaNi$_{2}$(PO$_{4}$)$_{2}$, Z. Anorg. Allg. Chem. \textbf{627}, 687 (2001).

\bibitem{Dwight1960} K. Dwight and N. Menyuk, Magnetic properties of Mn$_3$O$_4$ and the canted spin problem, Phys. Rev. \textbf{119}, 1470 (1960).

\bibitem{Kittel} C. Kittel, \emph{Introduction to Solid State Physics} 38 4th ed. (Wiley, New York, 1966).

\bibitem{Koteswararao2014} B. Koteswararao, R. Kumar, P. Khuntia, S. Bhowal, S. K. Panda, M. R. Rahman, A. V. Mahajan, I. Dasgupta, M. Baenitz, K. H. Kim, and F. C. Chou, Magnetic properties and heat capacity of the three-dimensional frustrated $S$ = 1/2 antiferromagnet PbCuTe$_2$O$_6$, Phys. Rev. B \textbf{90}, 035141 (2014).

\bibitem{Yogi2015} A. Yogi, N. Ahmed, R. Nath, A. A. Tsirlin, S. Kundu, A. V. Mahajan, J. Sichelschmidt, B. Roy, and Y. Furukawa, Antiferromagnetism of ${\mathrm{Zn}}_{2}\mathrm{VO}{{(\mathrm{PO}}_{4})}_{2}$ and the dilution with ${\mathrm{Ti}}^{4+}$, Phys. Rev. B \textbf{91}, 024413 (2015).

\bibitem{Johnston2011} D. C. Johnston, R. J. McQueeney, B. Lake, A. Honecker, M. E. Zhitomirsky, R. Nath, Y. Furukawa, V. P. Antropov, and Y. Singh, Magnetic exchange interactions in BaMn$_2$As$_2$: A case study of the $J_1-J_2-J_c$ Heisenberg model, Phys. Rev. B \textbf{84}, 094445 (2011).

\bibitem{note3} A peak can be still seen in the magnetic susceptibility slightly above $T_N$, but this feature is common to Heisenberg antiferromagnets. Note also that the $T_{\rm max}/T_N$ ratio for the magnetic susceptibility is 1.3 in Ba$_2$Mn(PO$_4)_2$ vs. 1.6 in Ba$_2$Ni(PO$_4)_2$, thus confirming the much stronger tendency of the Ni compound to the short-range order above $T_N$.

\bibitem{Bertaut.JAP.33.1138} E. F. Bertaut, A. Delapalme, F. Forrat, and G. Roult, Magnetic Structure Work at the Nuclear Center of Grenoble, J. Appl. Phys. \textbf{33}, 1123 (1962).

\bibitem{Bertaut.ACa.24.217} E. F. Bertaut, Representation analysis of magnetic structures, Acta Cryst. \textbf{24}, 217 (1968).

\bibitem{Bertaut.JPC.32.C1-462} E. F. Bertaut, MAGNETIC STRUCTURE ANALYSIS AND GROUP THEORY, J. Phys. Colloques \textbf{32}, C1-462 (1971).

\bibitem{Bertaut.JMMM.24.267} E. F. Bertaut, On group theoretical techniques in magnetic structure analysis, J. Magn. Magn. Mat. \textbf{24}, 267 (1981).

\bibitem [{\citenamefont {Izyumov}\ \emph {et~al.}(1991)\citenamefont
  {Izyumov}, \citenamefont {Naish},\ and\ \citenamefont
  {Ozerov}}]{Izyumov-book}%
  \BibitemOpen
  \bibfield  {author} {\bibinfo {author} {\bibfnamefont {Y.~A.}\ \bibnamefont
  {Izyumov}}, \bibinfo {author} {\bibfnamefont {V.~E.}\ \bibnamefont {Naish}},
  \ and\ \bibinfo {author} {\bibfnamefont {R.~P.}\ \bibnamefont {Ozerov}},\
  }\href@noop {} {\emph {\bibinfo {title} {Neutron Diffraction of Magnetic
  Materials}}}\ (\bibinfo  {publisher} {Consultants Bureau, New York, USA},\
  \bibinfo {year} {1991}).

\bibitem [{\citenamefont {Bradley}\ and\ \citenamefont
  {Cracknell}(1972)}]{Bradley_Cracknell}%
  \BibitemOpen
  \bibfield  {author} {\bibinfo {author} {\bibfnamefont {C.}~\bibnamefont
  {Bradley}}\ and\ \bibinfo {author} {\bibfnamefont {A.}~\bibnamefont
  {Cracknell}},\ }\href@noop {} {\emph {\bibinfo {title} {The Mathematical
  Theory of Symmetry in Solids}}}\ (\bibinfo  {publisher} {Clarendon Press,
  Oxford},\ \bibinfo {year} {1972}).

\bibitem [{\citenamefont {Cracknell}(1975)}]{Cracknell}%
  \BibitemOpen
  \bibfield  {author} {\bibinfo {author} {\bibfnamefont {A.}~\bibnamefont
  {Cracknell}},\ }\href@noop {} {\emph {\bibinfo {title} {Magnetism in
  Crystalline Materials}}}\ (\bibinfo  {publisher} {Pergamon Press, Oxford},\
  \bibinfo {year} {1975}).

\bibitem{Wills.PRB.63.064430} A.~S. Wills, Long-range ordering and representational analysis of the jarosites, Phys. Rev. B \textbf{63}, 064430 (2001).

\bibitem{note1} Note that the $\Gamma_1$ magnetic structure of \BNPO\ is mistakenly identified as $\Gamma_2$ in Ref.~\onlinecite{Yogi2017}, see also Phys. Rev. B \textbf{96}, 059903 (2017).

\bibitem{note2} The fitting parameters are $g=1.97$ and $J_3=1.52$\,K in good agreement with the DFT results from Table~\ref{tab:exchange}.

\bibitem [{\citenamefont {Momma}\ and\ \citenamefont {Izumi}(2011)}]{vesta}%
  \BibitemOpen
  \bibfield  {author} {\bibinfo {author} {\bibfnamefont {K.}~\bibnamefont
  {Momma}}\ and\ \bibinfo {author} {\bibfnamefont {F.}~\bibnamefont {Izumi}},\
  }\href@noop {} {\bibfield  {journal} {\bibinfo  {journal} {J. Appl.
  Crystallogr.}\ }\textbf {\bibinfo {volume} {44}},\ \bibinfo {pages} {1272}
  (\bibinfo {year} {2011})}.


\end{thebibliography}
\end{document}